\documentclass[useAMS,usenatbib]{mnras}
\usepackage{amsmath,psfig,epsfig,graphics}
\usepackage{graphicx}
\usepackage{epstopdf}
\usepackage{float}
\usepackage{enumerate}
\usepackage{natbib}
\usepackage{threeparttable}
\usepackage{xcolor}

\def\chandra    {{\em Chandra}\/}

\def\hst   {{\em HST}\/}

\def\rosat      {{\em ROSAT}\/}

\def\arcdeg{\hbox{$^\circ$}}
\def\arcmin{\hbox{$^\prime$}}
\def\arcsec{\hbox{$^{\prime\prime}$}}

\newcommand{\RNum}[1]{\uppercase\expandafter{\romannumeral #1\relax}}

\hypersetup{draft}
\begin{document}

\title[AGN feedback in 3C 220.1]{AGN feedback in the FR II galaxy 3C 220.1}

\author[Liu et al.]
{Wenhao Liu${}^1$\thanks{E-mail: wl0014@uah.edu}, Ming Sun${}^1$\thanks{E-mail: ms0071@uah.edu},
Paul E. J. Nulsen${}^{2,3}$, Diana M. Worrall${}^4$, Mark Birkinshaw${}^4$, \newauthor
Craig Sarazin${}^5$, William R. Forman${}^2$, Christine Jones${}^2$, Chong Ge${}^1$\\
\\
$^{1}$ Department of Physics and Astronomy, University of Alabama in Huntsville, Huntsville, AL 35899, USA\\
${}^2$ Harvard-Smithsonian Center for Astrophysics, 60 Garden Street, Cambridge, MA 02138, USA \\
${}^3$ ICRAR, University of Western Australia, 35 Stirling Hwy, Crawley, WA 6009, Australia\\
${}^4$ HH Wills Physics Laboratory, University of Bristol, Tyndall Avenue, Bristol BS8 1TL, UK \\
${}^5$ Department of Astronomy, University of Virginia, Charlottesville, VA 22904, USA\\
}


\maketitle
\label{firstpage}

\begin{abstract}
	We present results from a deep (174 ks) \chandra\ observation of the FR-II radio galaxy 3C~220.1, the central 
	brightest cluster galaxy (BCG) of a $kT \sim$ 4 keV cluster at $z=0.61$. 
	The temperature of the hot cluster medium drops from $\sim5.9$ keV to $\sim3.9$ keV at $\sim$ 35 kpc radius, while the temperature at smaller radii may be substantially lower. 
	The central active galactic nucleus (AGN) outshines the whole cluster in X-rays, with a bolometric luminosity of $2.0\times10^{46}$ erg s$^{-1}$ ($\sim10$\% of the Eddington rate).
	The system shows a pair of potential X-ray cavities $\sim35$ kpc east and west of the nucleus. 
	The cavity power is estimated within the range of $1.0\times10^{44}$ erg s$^{-1}$ and $1.7\times10^{45}$ erg s$^{-1}$, from different methods.
	The X-ray enhancements in the radio lobes could be due to inverse Compton emission, with a total 2-10 keV 
	luminosity of $\sim8.0\times10^{42}$ erg s$^{-1}$. We compare 3C~220.1 with other cluster BCGs, including Cygnus~A, 
	as there are few BCGs in rich clusters hosting an FR-II galaxy. We also summarize the jet power of FR-II galaxies from different methods. The comparison suggests that the cavity power of FR-II galaxies likely under-estimates the jet power. The properties of 3C~220.1 suggest that it is 
	at the transition stage from quasar-mode feedback to radio-mode feedback.
\end{abstract}
\begin{keywords}
galaxies: groups: individual: 3C 220.1 -- X-rays: galaxies: clusters -- galaxies: jets
\end{keywords}
\section{Introduction}
\label{intro}
It has been widely accepted that feedback from central active galactic nuclei (AGN) plays an important
role in galaxy formation and evolution \citep[e.g.,][]{McNamara07,Fabian12}. 
In the nearby Universe, a strong central AGN, hosted by the brightest cluster galaxy (BCG) in the core of 
the galaxy cluster, can drive energetic radio jets, which propagate through the intracluster medium (ICM), 
push aside the hot X-ray gas, and create cavities visible in X-ray images \citep[e.g.,][]{Churazov01}. 
At low redshift, the X-ray cavities are predominantly found in cool-core clusters, with a high detection rate of 
90\% \citep[e.g.,][]{Dunn06,Fabian12}, although cases in non-cool core clusters have also been reported
\citep[e.g.,][]{Worrall17}. This is known as ``radio-mode'' AGN feedback, which is mainly mechanically 
dominated and occurs when the central black holes (BHs) are accreting at a low rate compared to the 
Eddington limit. The radio AGN affects the ICM through AGN-driven shocks and bubbles.
The high mechanical power of radio AGN can not only quench cooling in cluster cool cores, but can also 
drive the ICM properties away from those defined by simple self-similar relations involving only gravity 
\citep[e.g.,][]{Voit05}. The features in the ICM, such as cavities and shocks discovered by high-resolution 
X-ray observations, can serve as calorimeters for the total energy output of the central AGN, allowing 
important constraints on the evolution of the central super-massive black hole (SMBH) \citep[e.g.,][]{McNamara07,Fabian12}.

Large cavities in the ICM and the associated AGN shocks trace the biggest explosions in the Universe. 
A correlation between the AGN radio luminosity and jet power, or the $L_{radio} - P_{jet}$ relation, 
has been established \citep{Birzan08,Cavagnolo10,OSullivan11}.
Thus, a natural approach to study the biggest AGN explosions is to study
the most powerful FR II galaxies. However, very few powerful FR II galaxies have been 
included in the studies of X-ray cavities. 
In the combined samples from the literature \citep[e.g.,][]{Birzan08,Cavagnolo10,OSullivan11,Hlavacek-Larrondo12},
Cygnus A and 3C 295 are the only two FR II galaxies. Besides these, the powerful
FR II radio galaxies, 3C 444 \citep{Croston11,Vagshette17}, and 3C 320 \citep{Vagshette19},
have been studied in detail for X-ray cavities.

At high redshift ($z>1$), AGN feedback is mainly radiatively dominated, in the so-called ``quasar-mode'',
which occurs when the central BHs are accreting at rates near the Eddington limit. Using a sample
of $z>0.3$ BCGs, \citet{Hlavacek-Larrondo12,Larrondo13} 
found that the X-ray AGN luminosities in BCGs increase with redshift, while the mechanical properties 
of the AGN outflows remain unchanged. The central AGN in such systems provide both strong radiative 
and mechanical feedback, suggesting that they are in the process of transitioning between quasar-mode 
feedback and radio-mode feedback \citep[e.g.,][]{OSullivan12,Larrondo15}.
The detection rate of cavities in BCGs at higher redshift is relatively small, e.g., $\sim25$\% in a sample of
76 clusters in Massive Cluster Survey within $0.3<z<0.7$ \citep{Hlavacek-Larrondo12} and $\sim7$\% in a sample 
of 86 SPT-selected clusters within $0.4<z<1.2$ \citep{Larrondo15}, as in general, 
long \chandra\ observations are required to detect the cavities in X-rays at high redshift.
The systems showing both radiative feedback and radio-mode feedback with detailed X-ray studies are very few.
These include H1821+643 ($z=0.299$) \citep{Russell10}, IRAS 09104+4109 ($z=0.442$) \citep{OSullivan12},
and the Phoenix cluster ($z=0.596$) \citep{McDonald15}.
It is crucial to build a census of AGN feedback in such high redshift systems to
understand how and why the SMBHs switch from one mode to the other.

3C~220.1 is classified as an FR II narrow emission line radio galaxy at redshift $z=0.61$ \citep{Spinrad85}, 
with two prominent radio lobes that extend over 100 kpc from the nucleus.
In Fig. \ref{radio} we show the VLA images of 3C~220.1 at 1.5, 4.9 GHz from program AA267, 
and 8.4 GHz from program AP380 \citep[e.g.,][]{Worrall01,Mullin06,Fernini14}.
The radio source has a core, one-sided deflected radio jet to the east and three ``hotspot-like'' 
structures as shown in Fig. \ref{radio} bottom right.
3C~220.1 was observed with the \rosat\ HRI, which allowed the emission to be decomposed into
a compact core and an extended component \citep[e.g.,][]{Hardcastle98,Worrall01,Belsole07}.
A subsequent short ($\sim18$ ks) \chandra\ observation confirmed two components, and showed that 3C~220.1
is the BCG of a massive cluster with a temperature of $\sim5$ keV \citep{Worrall01}.
Although optical observations show no evidence of a rich cluster near 3C~220.1, the galaxy cluster 
containing 3C~220.1 hosts a giant luminous arc, which is $\sim9$\arcsec\ away from the center and subtends 
$\sim70$\arcdeg\ around the radio galaxy, as observed with the {\em Hubble Space Telescope} (\hst) in Fig. \ref{hst_image}.
The cluster is indeed massive for its redshift, with a virial mass
of $3.5\times10^{14}$ $M_{\odot}$ estimated from strong lensing \citep{Comerford07}.
The cluster hosts a large X-ray cool core, with a central cooling time less than 1 Gyr \citep{Worrall01}.
The BCG is still actively forming stars, with \hst\ data suggesting a star formation rate 
of $\sim21- 67 $ $M_{\odot}$ yr$^{-1}$ \citep{Westhues16}. 
We summarize the cluster properties of 3C~220.1 in Table \ref{tab_info}. 
The central X-ray AGN is very luminous with a 0.7-12 keV luminosity
of $\sim10^{45}$ erg s$^{-1}$ \citep{Worrall01}, suggesting that 3C~220.1 is undergoing strong radiative feedback. 
Here, we present a detailed study with deep (170 ks) \chandra\ observations to explore the AGN feedback in 3C~220.1.

In this paper we adopt a cosmology with $H_{\rm 0}=70$ km s$^{-1}$ Mpc$^{-1}$, $\Omega_{\rm M}=0.3$,
and $\Omega_{\rm \Lambda}=0.7$. At a redshift of $z=0.61$, the luminosity distance is 3601.4 Mpc and 
$1\arcsec$ corresponds to 6.736 kpc. All error bars are quoted at $1\sigma$ confidence level, unless otherwise specified.
\begin{table*}
\protect\caption{Properties of the 3C~220.1 cluster}
\begin{tabular}{|c|c|c|c|c|c|c|c|}
\hline 
	Name & $z$ & $kT_{500}$$^{a}$    & Z$^{a}$ & $L_{X,500}$$^{a}$ & M$_{500}$$^{a}$ & P$_{\rm 1.4GHz}$  & radio spectral index$^b$ \\
	&   & (keV)         &           & ($10^{44}$ erg s$^{-1}$) & ($10^{14}$ M$_{\odot}$) & ($10^{27}$ W Hz$^{-1}$)  &  \tabularnewline
\hline 
	3C~220.1 & 0.61 & $3.7_{-0.6}^{+0.9}$ & $0.5_{-0.4}^{+0.6}$ & $2.5\pm0.4$  &  $1.3_{-0.3}^{+0.6}$ & $3.4\pm0.1$ & $-0.97\pm0.02$ \tabularnewline
\hline 
\end{tabular}
\begin{tablenotes}
\item
 $a$: Temperature and abundance are measured within $0.15-0.75 R_{500}$ ($\sim16\arcsec-80\arcsec$). 
	The X-ray bolometric luminosity is measured within $R_{500}$ based on the surface brightness profile. 
	The gravitational mass is estimated based on the $M-T$ relation \citep{Sun09}. Based on the established 
	$L-T$ relation, the estimated luminosity from the hot ICM, after correcting for the cosmological evolution 
	term $E(z)$, is $\sim2.0\times10^{44}$ erg s$^{-1}$ from \cite{Giles16}, or $\sim4.2\times10^{44}$ 
	erg s$^{-1}$ from \cite{Sun12}.
	$b$: The fitted radio spectral index across the whole source from 38 MHz to 10 GHz based on the data from NED. \\
\end{tablenotes}
\label{tab_info}
\end{table*}

\section{Chandra Data Analysis}
\label{DataAnalysis}
\begin{table}
\protect\caption{\chandra\ Observations of 3C~220.1 (PI: Sun)}
\begin{tabular}{|c|c|c|c|}
\hline 
ObsID & Date Obs & Total Exp    & Cleaned Exp \\
      &          & (ks)         & (ks)        \tabularnewline
\hline 
17130 & 2016 Jan 17 & 41.51 & 40.94 \tabularnewline
17131 & 2016 Jun 02 & 49.40 & 49.13 \tabularnewline
18747 & 2016 Jan 25 & 54.34 & 54.25 \tabularnewline
18860 & 2016 Jun 05 & 28.65 & 28.65 \tabularnewline
\hline 
\end{tabular}
\label{tab_obs}
\end{table}
3C 220.1 was observed with \chandra\ for $\sim170$ ks split into four observations 
(Obs. IDs 17130, 17131, 18747, and 18860, PI: Sun) in January and June 2016 with the 
Advanced CCD Imaging Spectrometer (ACIS). All four observations placed the core on chip S3 
and were taken in Very Faint (VFAINT) mode. The details of the \chandra\ observations are 
summarized in Table \ref{tab_obs}. There also was one $\sim$ 19 ks ACIS-S observation in 1999, 
published by \citet{Worrall01}. In this study we focus on the new, deep data in 2016, due 
to significant changes in the characteristics of the instrument between these dates.
The data were analyzed using CIAO 4.9 and CALDB version 4.7.4 from the \chandra\ X-ray Center. 
For each observation the level 1 event files were reprocessed using the
{\tt CHANDRA\_REPRO} script to account for afterglows, bad pixels, 
charge transfer inefficiency, and time-dependent gain correction. The improved background filtering 
was also applied by setting {\tt CHECK\_VF\_PHA} as ``yes'' to remove bad events that are likely 
associated with cosmic rays. The background light curve extracted from a source-free region was 
filtered with the {\tt LC\_CLEAN} script\footnote{http://asc.harvard.edu/contrib/maxim/acisbg/} to 
identify any period affected by background flares. There were no strong background flares for 
these observations and the resulting cleaned exposure time is given in Table \ref{tab_obs}.

In order to correct for small astrometric errors, we chose ObsID 18747 as the reference due to 
its longest exposure. For each observation, a broad band (0.5-7.0 keV energy band) image on the 
S3 chip was made and point sources were detected. The astrometric translation required to align 
each data set with the reference data was obtained using the CIAO tool
{\tt wcs\_match}, and was applied to each event list using the CIAO tool {\tt wcs\_update}. 

Point sources were detected in the combined 0.5-7.0 keV count image using the CIAO tool {\tt WAVDETECT}, 
with the variation of the point spread function across the field considered.
The detection threshold was set to $10^{-6}$ and the scales are from 1 to 16 pixels, increasing in
steps of a factor of $\sqrt{2}$. All detected sources were visually inspected and masked in the analysis 
except the central AGN.
A weighted exposure map was generated to account for quantum efficiency, vignetting
and the energy dependence of the effective area assuming an absorbed APEC model with $kT = 4.0$ keV,
$N_{\textrm{H}} = 2.0\times10^{20}$ cm$^{-2}$ (the weighted column density of the total Galactic 
hydrogen absorption, calculated by the ``NHtot'' tool\footnote{http://www.swift.ac.uk/analysis/nhtot/index.php}), 
and abundance of 0.3 $Z_{\odot}$ at the redshift $z = 0.61$.
We used the CALDB blank sky background files in our data analysis. For each observation the standard 
blank sky file for each chip was reprojected to match the time dependent aspect solution, and normalized 
to match the count rate in the 9.5-12.0 keV band to account for variations in the particle background. 
We estimated the local X-ray background for 3C~220.1 in each observation using a region 
where the surface brightness is approximately constant. The regions are $>3.0\arcmin$ from the cluster 
on the S3 chip. For spectral fitting, we used XSPEC version 12.10.0 and AtomDB 3.0.9, assuming the 
solar abundance table by \citet{aspl}. The absorption model is {\tt TBABS}.

\section{Spatial and Spectral Analysis}
\subsection{Spatial Analysis}
\begin{figure*}
\hbox{\hspace{-15px}
\begin{tabular}{ll}
\includegraphics[scale=0.37]{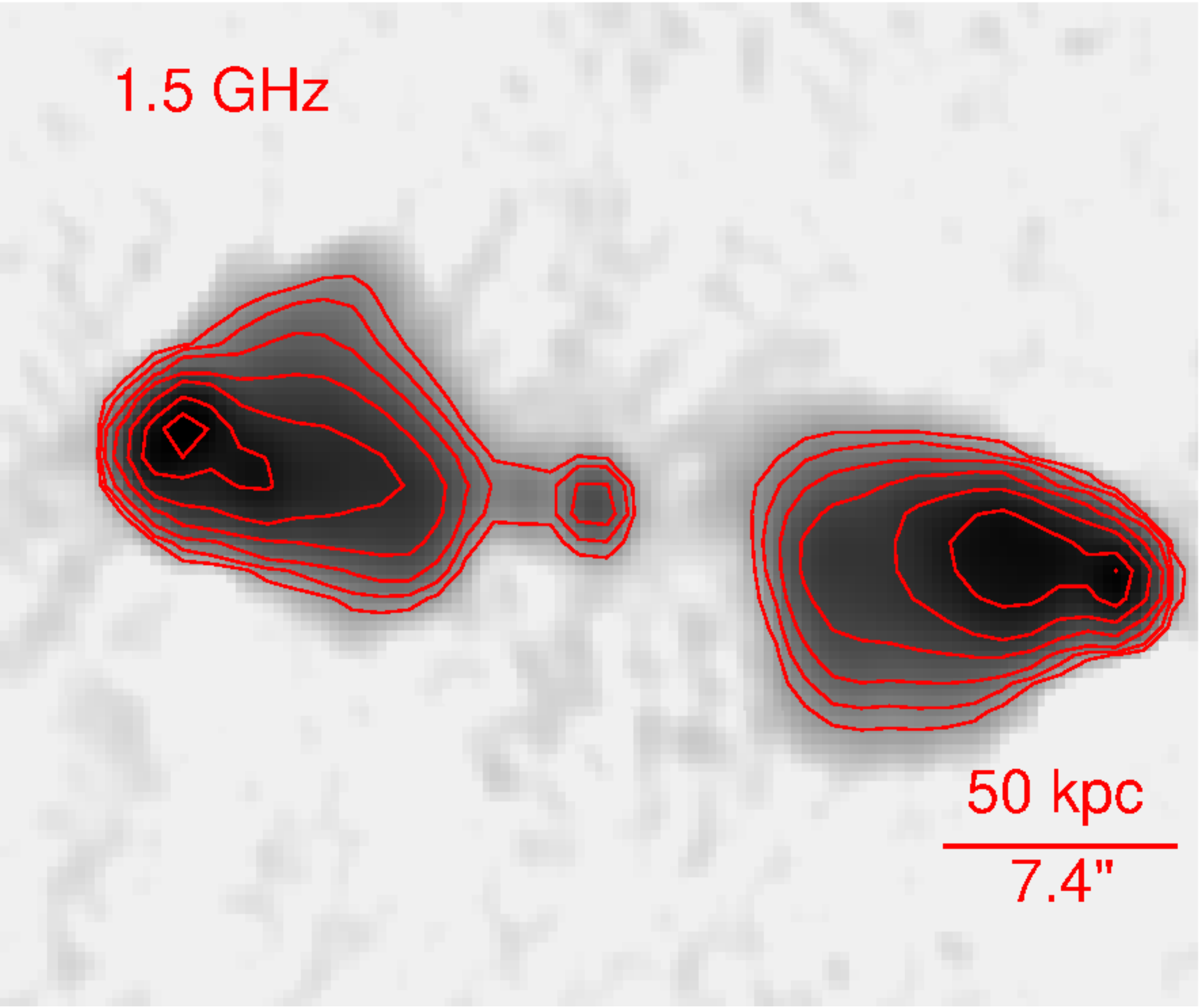}
&
\hspace{-10px}
\includegraphics[scale=0.37]{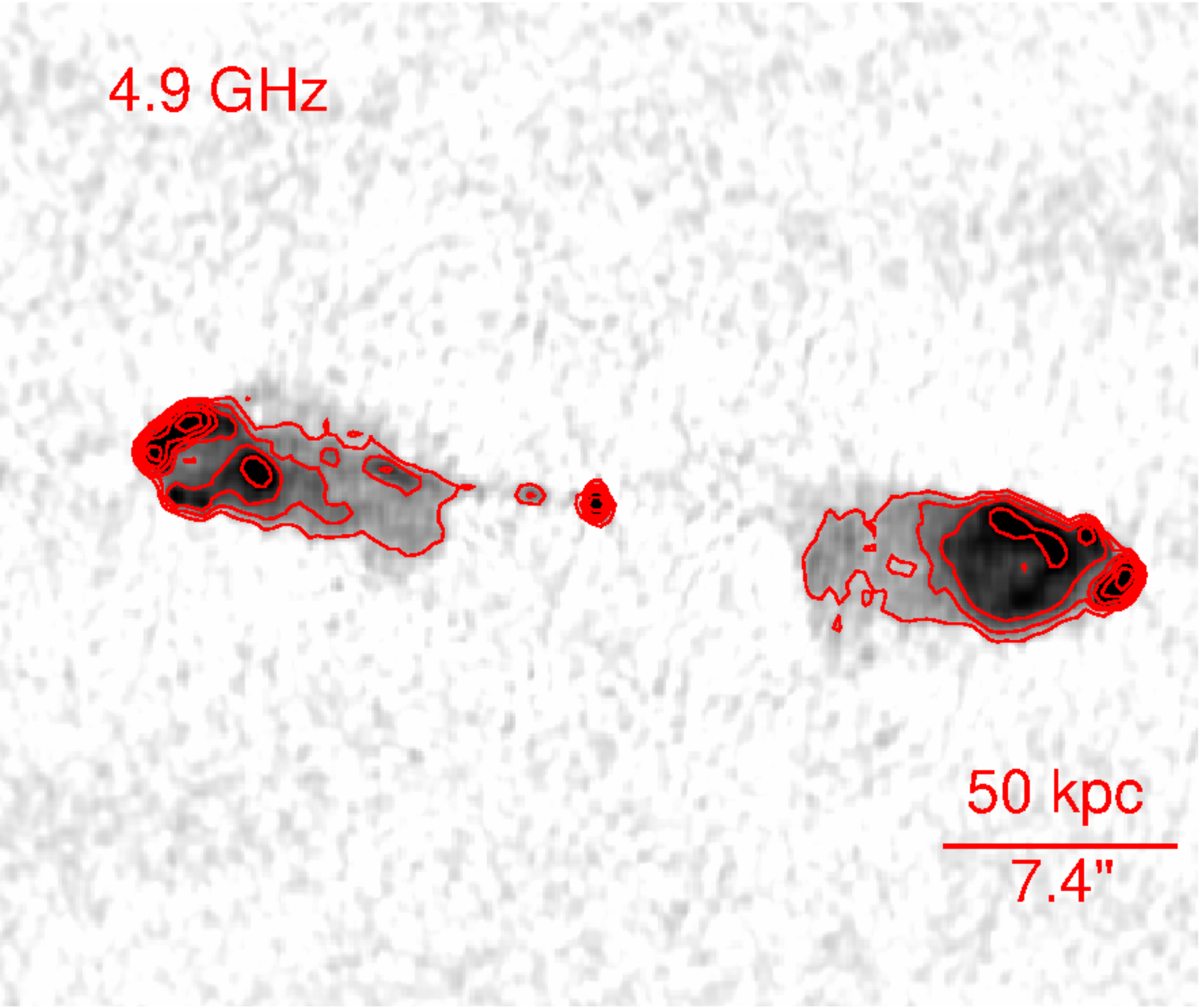}
\\
\includegraphics[scale=0.37]{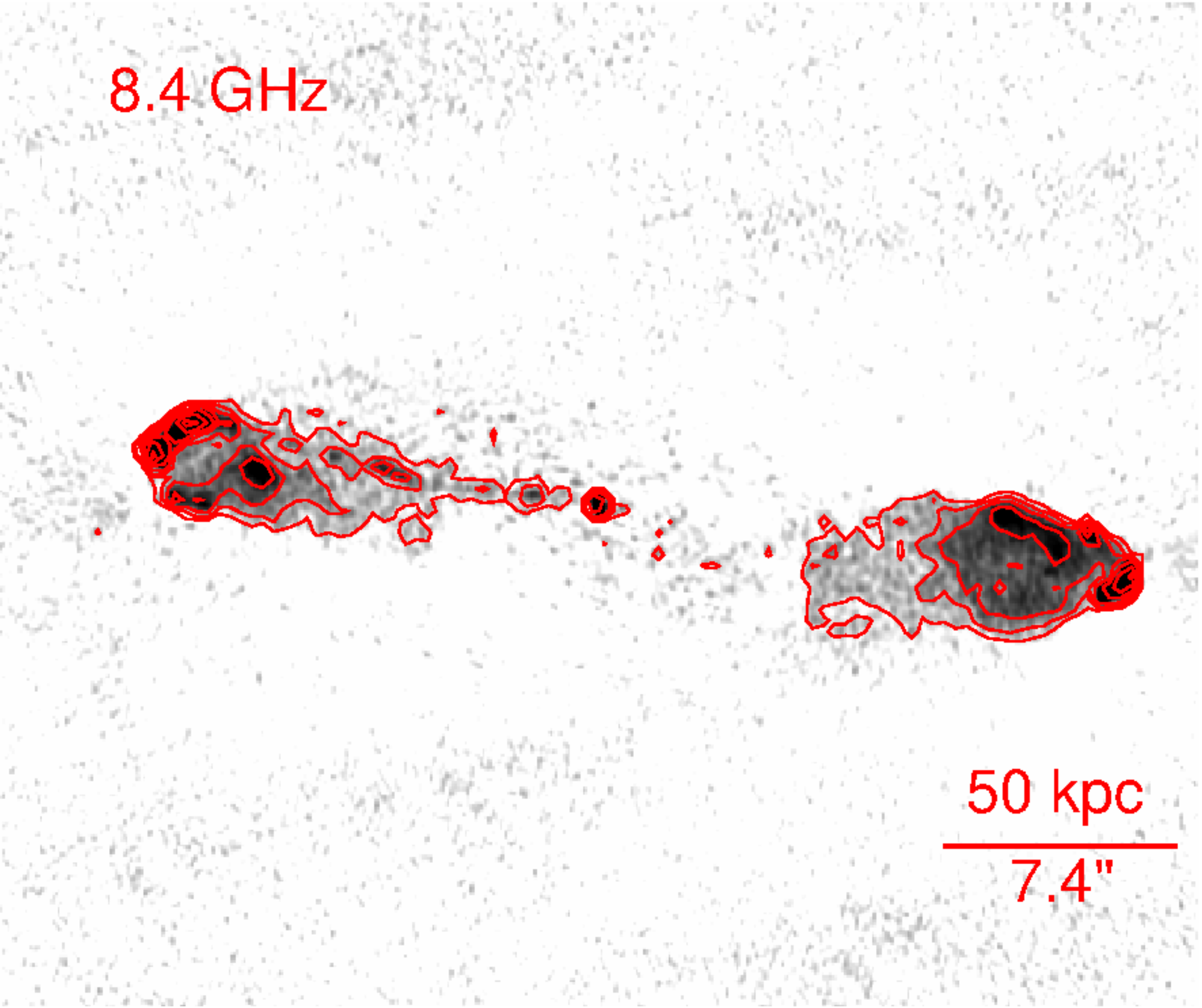}
&
\hspace{-10px}
\includegraphics[scale=0.37]{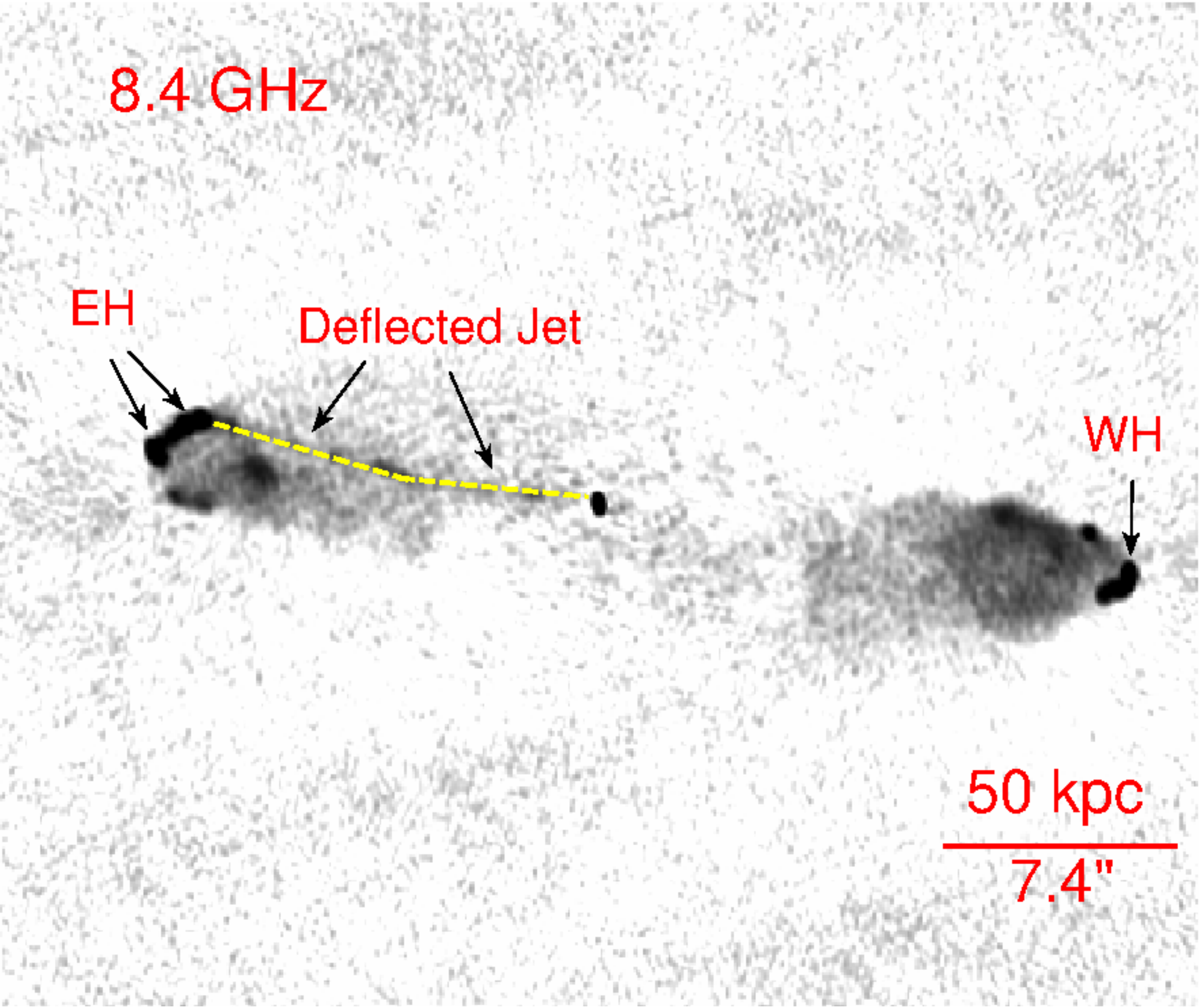}

\end{tabular}
}
\caption{ 
	VLA images and contours of 3C~220.1 at 1.5 GHz (top left), 4.9 GHz (top right), and 8.4 GHz 
	(bottom left) with resolution of $1.8\times1.4$ arcsec, $0.5\times0.3$ arcsec, and 
	$0.3\times0.2$ arcsec. The contours start at 2$\sigma$ and are spaced by a factor of 2 in 
	intensity. The 8.4 GHz radio image (bottom right) shows three ``hotspot-like'' structures 
	(two eastern hotspots, EH, and one western hotspot, WH) and a deflected jet (with a deflection 
	angle of $\sim14\arcdeg$) on the eastern side (highlighted with dashed yellow curve). 
	The projected distances of the farthest points of the radio lobes to the nucleus are $\sim97$ 
	and $\sim120$ kpc for the east and west, respectively.
	}
\label{radio}
\end{figure*}

\begin{figure}
\hbox{\hspace{-4px}
\includegraphics[scale=0.35]{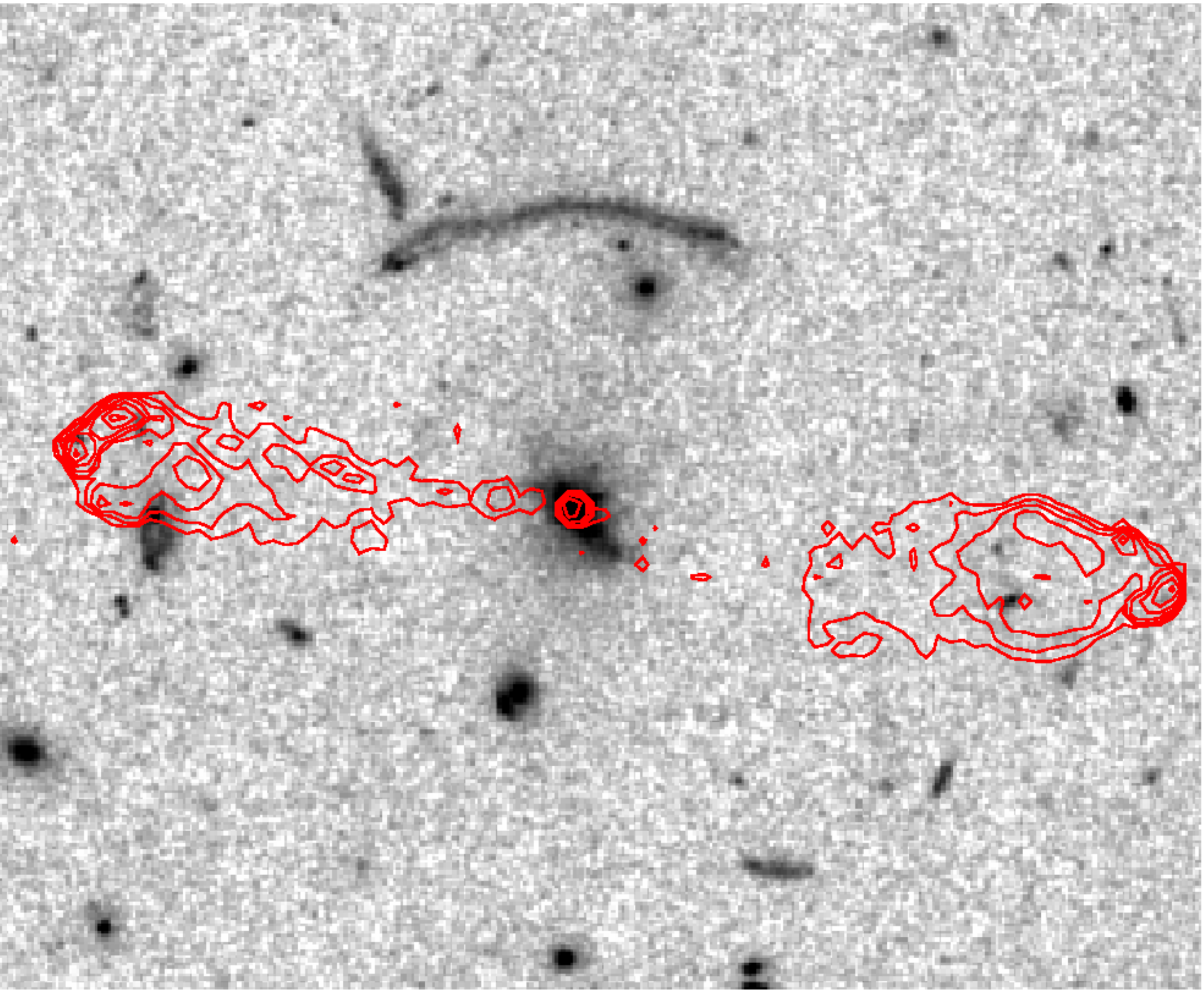}
}
\caption{
	The \hst\ WFPC2 F555W image of 3C~220.1 (roughly corresponding to the rest-frame $U$ band), 
	from the \hst\ program ID 6778, overlaid with the VLA 8.4 GHz radio contours (in red) as in 
	Fig. \ref{radio}. This shows the irregular morphology of the BCG that may imply star formation 
	and a large strong-lensing arc to the north.
}
\label{hst_image}
\end{figure}

\begin{figure*}
\hbox{\hspace{10px}
\includegraphics[scale=0.47]{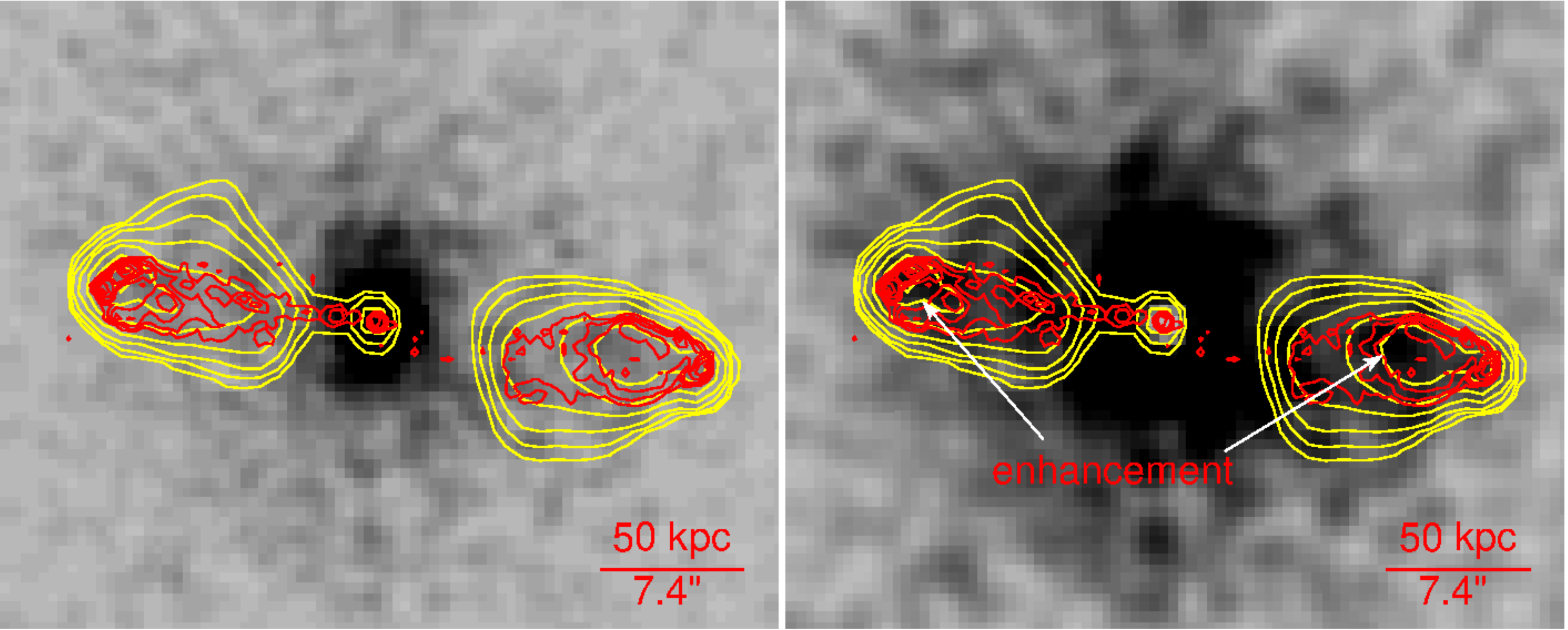}
}
\caption{
	{\em Left}: The combined background-subtracted, exposure-corrected \chandra\ image of 3C~220.1 
	in the 0.5-3.0 keV band, smoothed with a two dimensional Gaussian 2-pixel kernel (0.492$''$/pixel) 
	and overlaid with the VLA 1.5 GHz (yellow) and 8.4 GHz (red) radio contours. 
        {\em Right}: The same \chandra\ image as in the left panel, smoothed with a 2D Gaussian 3-pixel kernel 
	(0.492$''$/pixel). We have masked the central bright AGN (2\arcsec\ radius to cover much of the PSF).
	X-ray enhancements are visible in eastern and western radio lobe regions.
}
\label{image}
\end{figure*}

\begin{figure*}
\hbox{\hspace{5px}
\begin{tabular}{ll}
\hspace{-0.5cm}
\includegraphics[scale=0.4]{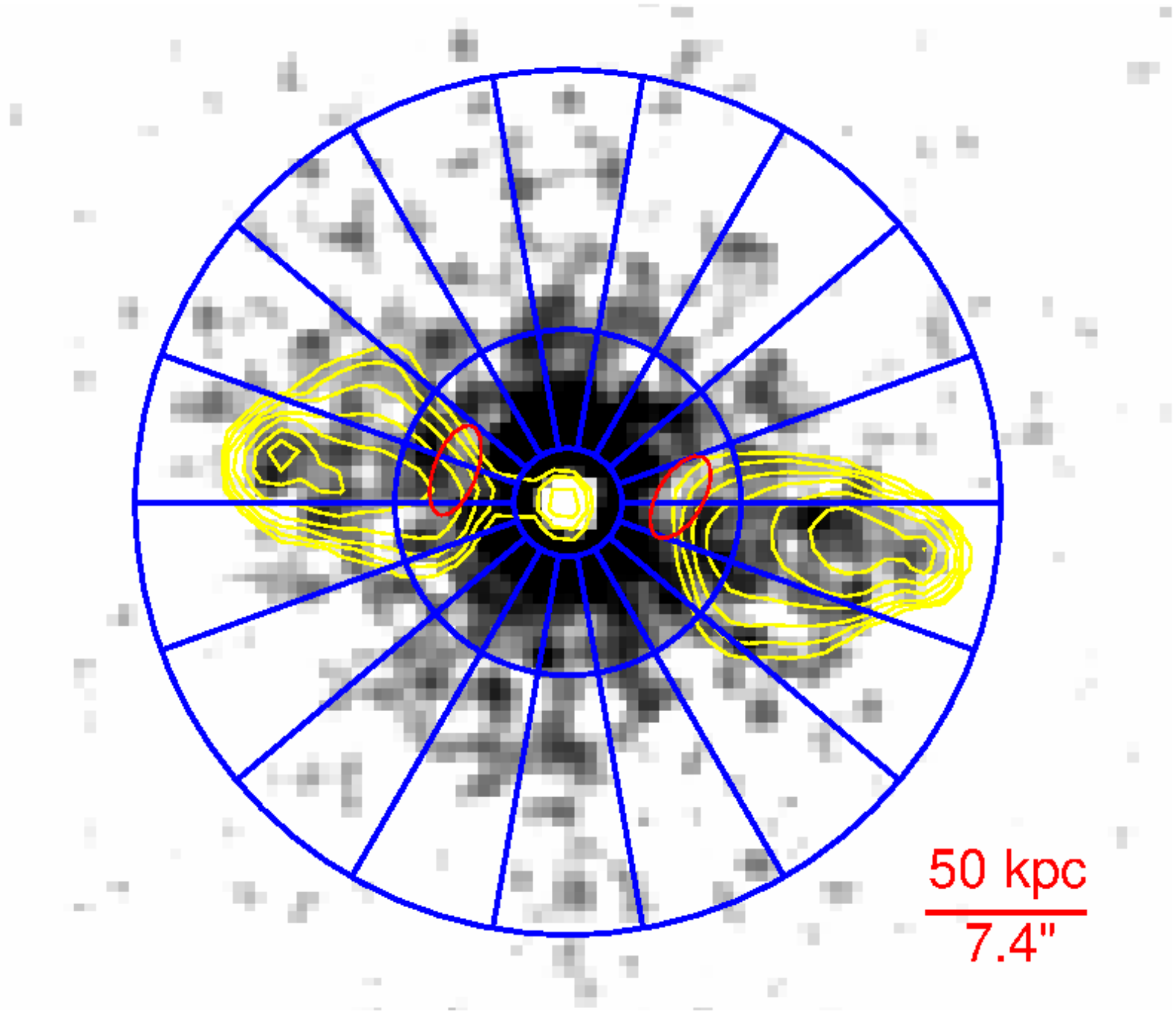}
&
\hspace{-1cm}
\includegraphics[scale=0.62]{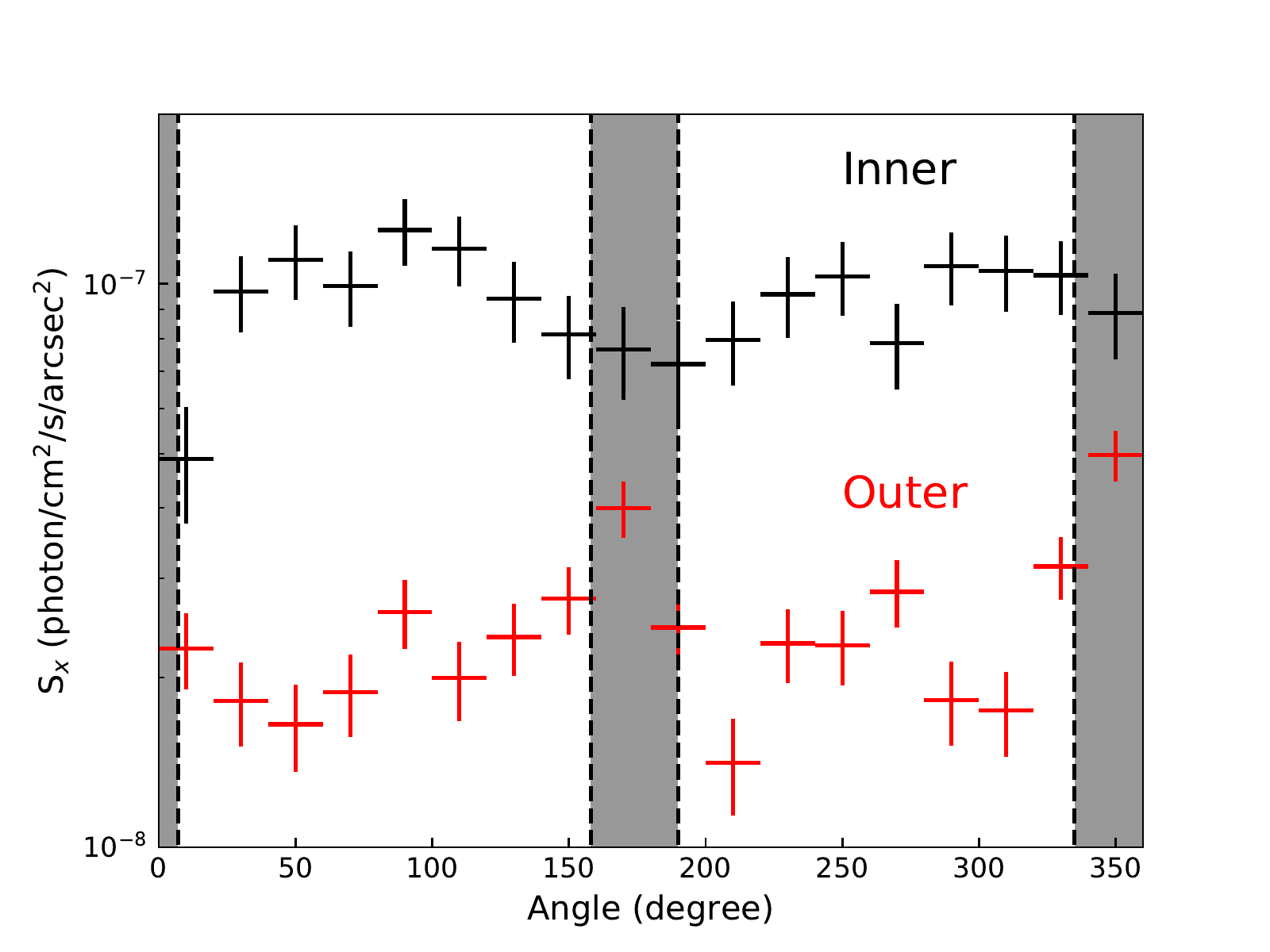}
\end{tabular}
}
\caption{
	Left: Exposure-corrected, background-subtracted \chandra\ image (0.5-2.0 keV) of 3C~220.1, smoothed 
	with a Gaussian 2-pixel kernel, overlaid with VLA 1.5 GHz contours. The overlaid blue regions denote 
	two annuli (3\arcsec-8\arcsec and 8\arcsec-20\arcsec) where the azimuthal surface brightness profiles 
	are extracted. The red ellipses denote the potential cavities showing surface brightness depressions 
	as in the right panel.
        Right: The surface brightness profiles extracted along the azimuthal directions (measured counterclockwise 
	from the west) with angle step size of 20 degrees from inner 3\arcsec-8\arcsec annulus (black) and outer 
	8\arcsec-20\arcsec annulus (red) in 0.5-2.0 keV band. The grey region between the dashed lines shows the 
	range for the eastern ($158\arcdeg-190\arcdeg$) and western ($-25\arcdeg-7\arcdeg$) radio lobes. The inner 
	surface brightness profile shows depressions around $-10\arcdeg-30\arcdeg$ and $150\arcdeg-210\arcdeg$. 
	The outer surface brightness profile shows enhanced emission in the region corresponding with the 
	eastern and western radio lobes.
}
\label{sb_in_out}
\end{figure*}

\begin{figure*}
\hbox{\hspace{5px}
\begin{tabular}{ll}
	\hspace{-1.2cm}
\includegraphics[scale=0.6]{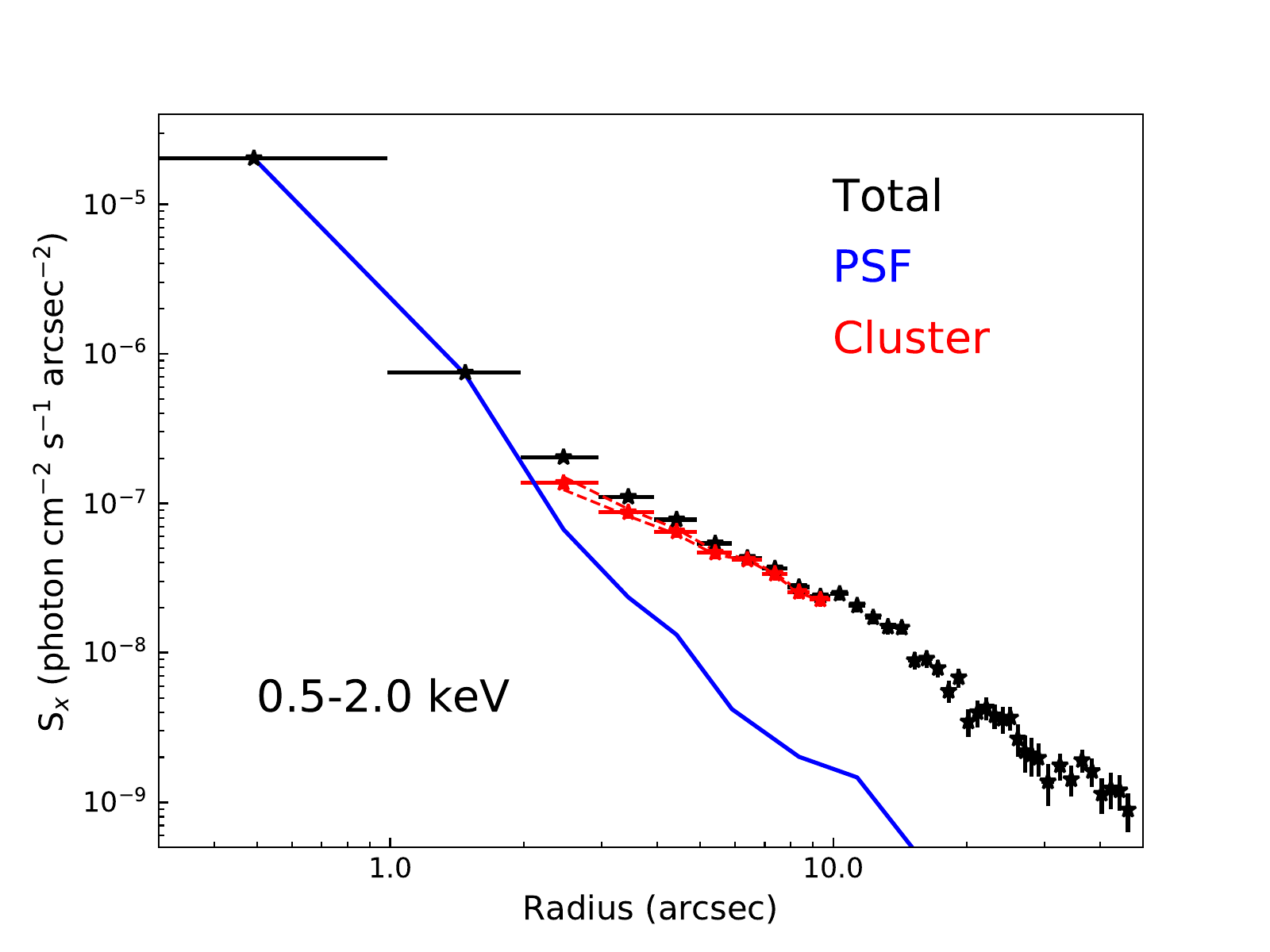}
&
\hspace{-1.2cm}
\includegraphics[scale=0.6]{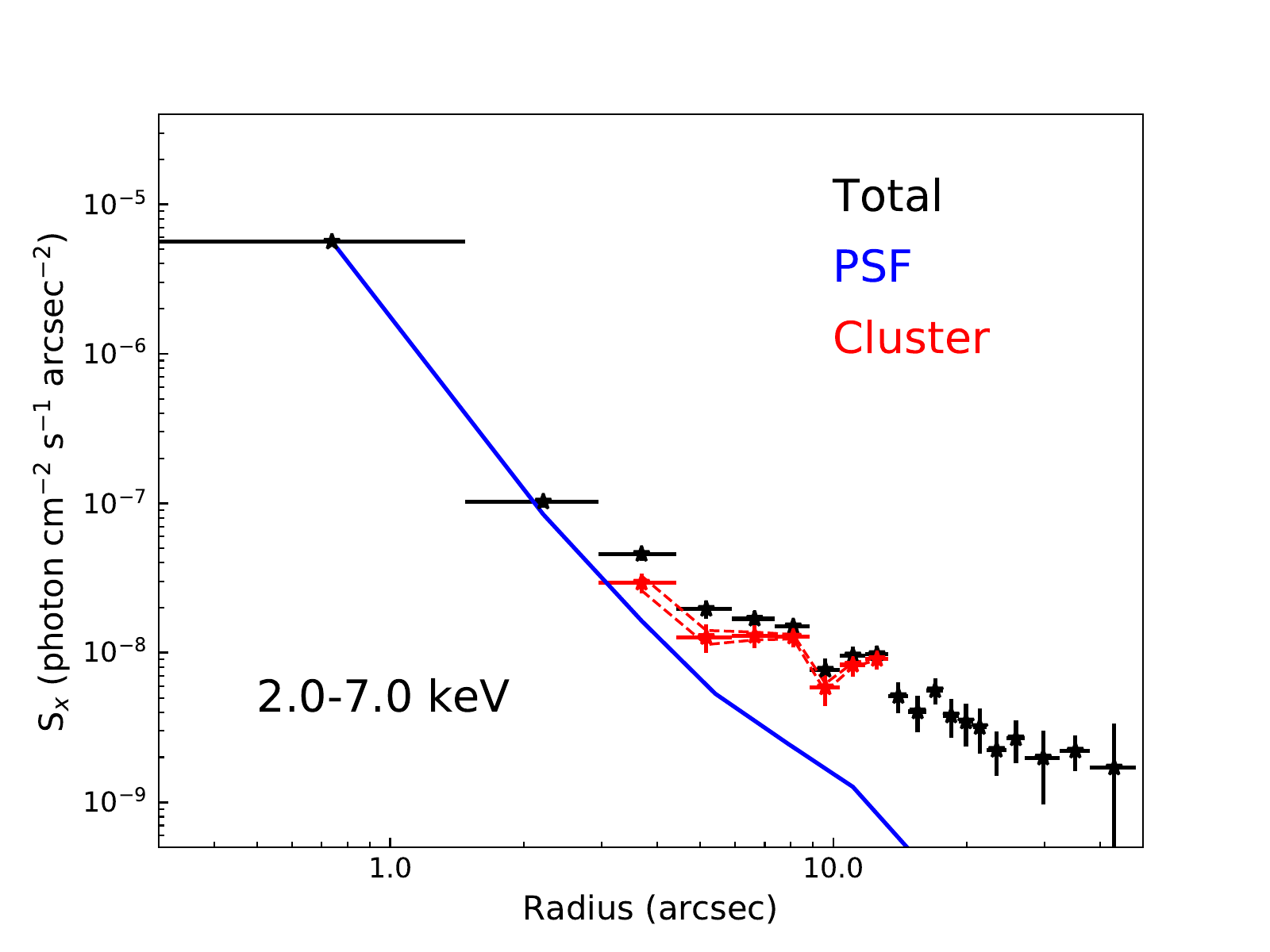}
\end{tabular}
}
\caption{
	Left: Total surface brightness profile in the energy range of 0.5-2.0 keV
	for 3C~220.1 (black), the simulated PSF (blue) of the central point source using ChaRT simulation,
	and the cluster surface brightness profile produced by subtracting the simulated PSF from the total 
	surface brightness (red). We normalized the PSF assuming that the central 2\arcsec\ is dominated
	by the AGN. The red dashed lines show the change on the cluster emission when the AGN normalization 
	is changed by 20\%.
	Right: Same as left, but in the 2.0-7.0 keV band.
	We do not plot the cluster surface brightness profile (red) to large radii since it would overlap
	with the total surface brightness profile in black.
}
\label{sbp_psf}
\end{figure*}

We reprojected and summed the count images, background images, and exposure maps respectively from four 
observations using the CIAO tool {\tt reproject\_image}. Fig. \ref{image} left panel shows the combined 
background-subtracted, exposure-corrected image in the 0.5-3.0 keV band, smoothed with a two dimensional 
Gaussian 2-pixel kernel ($0\arcsec.492$/pixel) and overlaid with the VLA 1.5 GHz and 8.4 GHz radio contours. 
Fig. \ref{image} right panel shows the same X-ray image, but with the central bright AGN masked, smoothed 
with a Gaussian 3-pixel kernel. X-ray enhancements, associated with the radio lobes, are visible in the figure.

Azimuthal surface brightness values were examined to determine the significance of potential cavities.
As shown in Fig. \ref{sb_in_out} (left panel), we extracted the surface brightness profiles in two 
annular regions centered on the nucleus with inner and outer radii of 
3\arcsec-8\arcsec\ and 8\arcsec-20\arcsec\ to investigate the azimuthal variations. Fig. \ref{sb_in_out} 
right shows the azimuthal variations of the surface brightness with an angle step size of 20\arcdeg\ in 
the inner and outer regions in the 0.5-2.0 keV  band. The inner surface brightness profiles show depressions
around 10\arcdeg-30\arcdeg\ and 150\arcdeg-210\arcdeg\ (measured counterclockwise from the west), because 
of the potential cavities, while the surface brightness profile in the outer region (corresponding to 
the radio lobes) shows enhanced emission. This may be due to inverse Compton X-ray emission 
(see Section \ref{Nonthermal}), which is often found in FR II radio lobes, and which was reported for 
3C~220.1 based on the earlier \chandra\ data by \citet{Croston05}.

\subsection{Cluster Surface Brightness Profile}
In order to study the arcsecond-features of the central bright source, we created subpixel (0.0984\arcsec/pixel) 
images by using the Energy-Dependent Subpixel Event Repositioning (EDSER) algorithm \citep{Li04}. 
The radial surface brightness profiles obtained from the subpixel images in the soft and hard X-ray 
bands for 3C~220.1 are shown in Fig. \ref{sbp_psf}. The central point source is bright and dominates 
the central region. To derive the gas properties for the cluster, we simulated the PSF of the central 
point source using the \chandra\ Ray Tracer (ChaRT) program \citep{Carter03}. We first fitted the central 
AGN spectrum with a power-law model (see section \ref{AGN}) and used the best-fitting model as the input 
for the ChaRT program. The simulation was then projected onto the ACIS-S detector with the MARX 
(version 5.3.2) software\footnote{https://space.mit.edu/CXC/MARX/} and used to estimate the PSF contamination 
from the total emission. The parameter {\tt AspectBlur} of MARX is used to account for the known 
uncertainty in the determination of the aspect solution. The parameter can be adjusted to better 
match the PSF core and wing to observations. In our study we used the default {\tt AspectBlur} 
value of $0\arcsec.25$ for ACIS-S observations. Fig. \ref{sbp_psf} shows the simulated PSF (blue) 
and the total observed surface brightness profile (black) in soft (0.5-2.0 keV) and hard (2.0-7.0 keV) 
bands. The PSF was normalized based on the total counts in the central 2\arcsec.

\subsection{Spectral Analysis}
\label{SpectralAnalysis}
\begin{figure*}
\begin{center}
\hbox{\hspace{5px}
\includegraphics[scale=1.0]{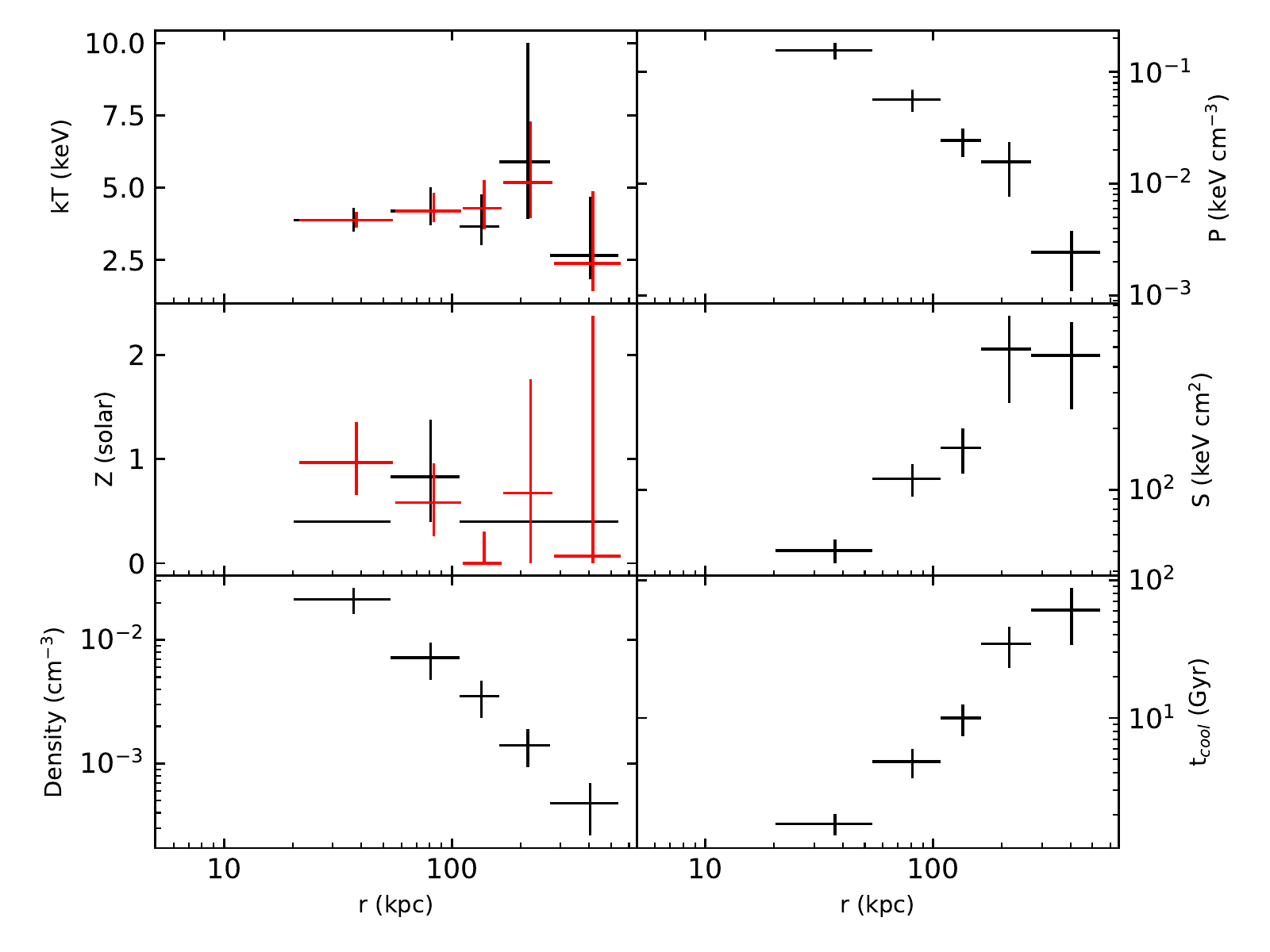}
}
\caption{\label{Tprofile}
	Radial profiles of the deprojected temperature, abundance, electron density, pressure, 
	entropy, and cooling time. We also show the projected temperature and abundance profiles in 
	red (we slightly shifted the red points horizontally for comparison). 
	We fix the deprojected abundance in all radial bins except the second bin.
	}
\end{center}
\end{figure*}

\begin{figure}
\begin{center}
\hbox{\hspace{5px}
\includegraphics[scale=0.55]{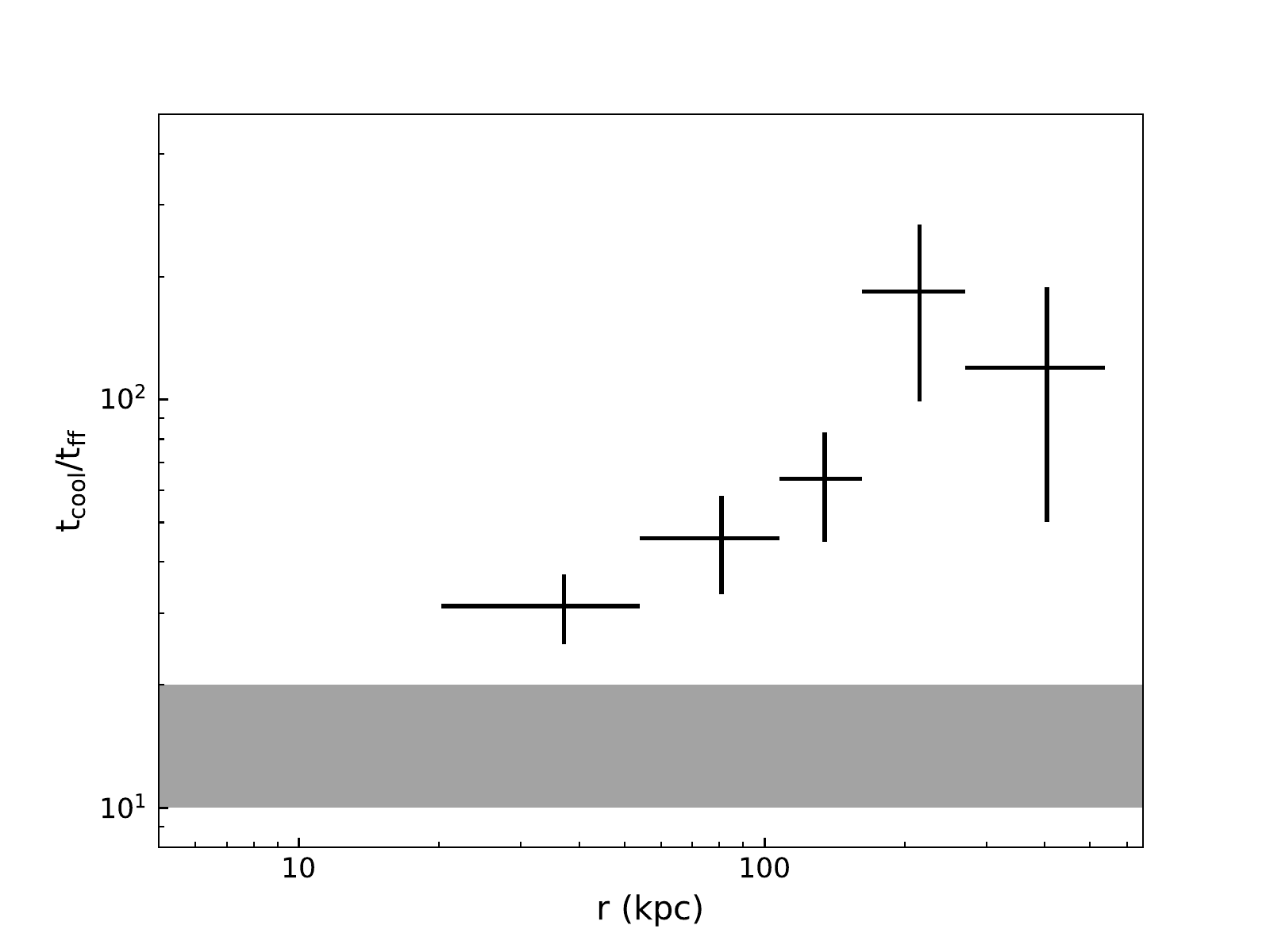}
}
\caption{\label{tratio}
	Radial profile of $t_{\rm cool}/t_{\rm ff}$. The grey region shows the typical threshold 
	for precipitation, min($t_{\rm cool}/t_{\rm ff}$)$\approx 10-20$, below which the hot gas 
	becomes thermally unstable \citep{Sharma12,Gaspari12a,Voit15a}.
	}
\end{center}
\end{figure}

\begin{figure}
\begin{center}
\hbox{\hspace{5px}
\includegraphics[scale=0.55]{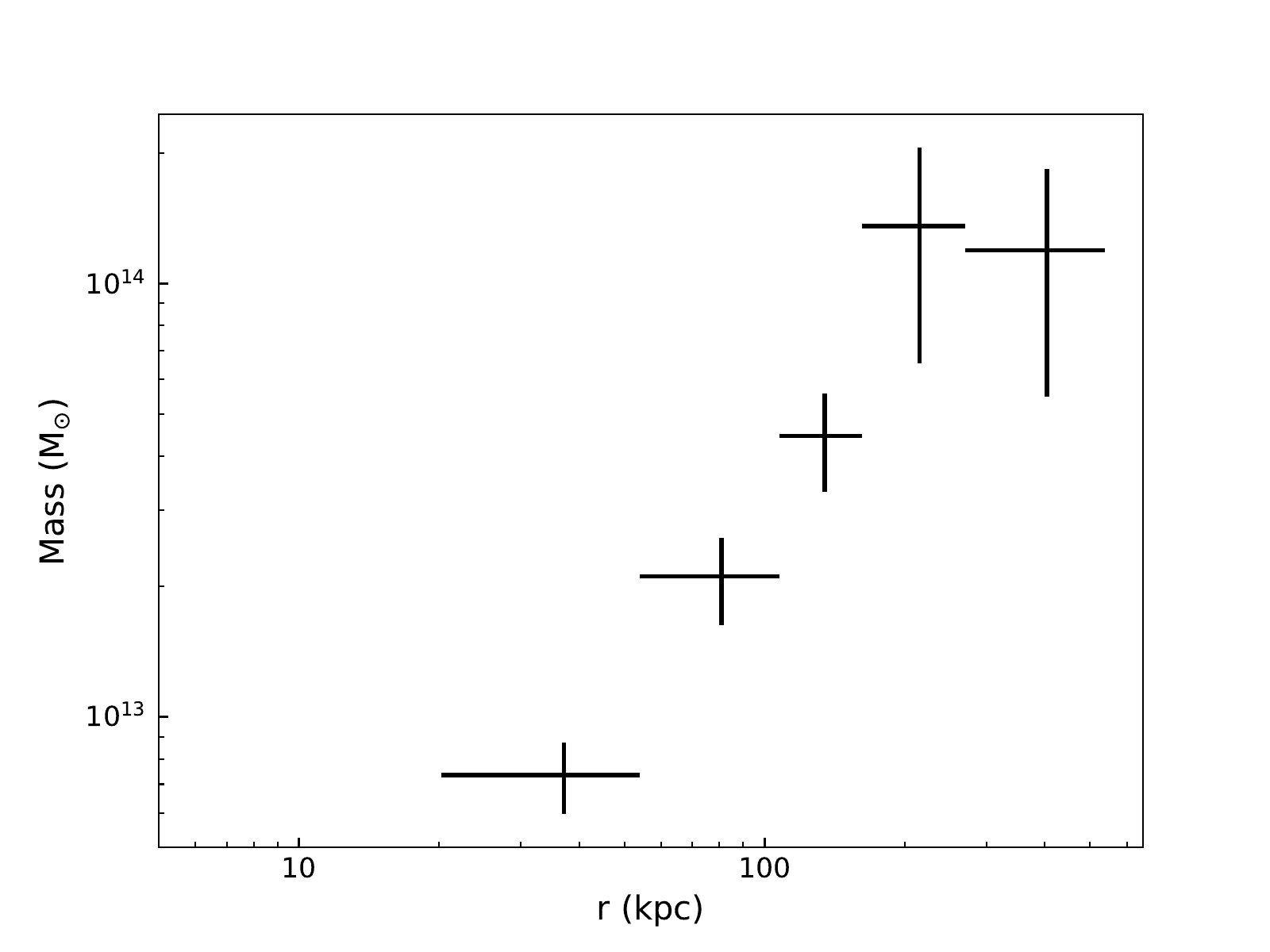}
}
\caption{\label{mass}
	The total gravitational mass profile for the cluster 3C~220.1 derived assuming hydrostatic 
	equilibrium and spherical symmetry.
}
\end{center}
\end{figure}

Spectra were extracted from the same circular annuli for the four observations separately using the 
CIAO tool {\tt specextract}. For each spectrum, the weighted response files and matrices were made, 
and a corresponding background spectrum was extracted from the same region from the blank sky background data. 
The spectra were fitted in the energy band 0.5-5.0 keV using the C-statistic \citep{Cash79}, since
it is largely unbiased compared to $\chi^{2}$ minimization \citep[e.g.,][]{Humphrey09}.
The spectra from the same region from four observations were fitted simultaneously to an absorbed thermal APEC model, i.e., {\tt TBABS*APEC}.
The temperature, metallicity, and the normalization were allowed to vary freely. For regions where 
the abundances cannot be well constrained due to the few source counts, we fixed the abundance at 
the value obtained in nearby regions.

We focused our analysis on the ACIS-S3 chip and extracted a set of 5 circular annuli from the center 
to a radius of $\sim540$ kpc. For the temperature deprojection, we adopted a model-independent approach 
\citep{Sun03,Liu19} which can account for point sources and chip gaps. The deprojected temperature 
profile is shown in Fig. \ref{Tprofile} (top left). We excluded the central $\sim20$ kpc radius 
($\sim3\arcsec$) due to the central bright point source. The deprojected temperature
decreases towards the center, from $\sim5.9$ keV at 215 kpc to $\sim3.9$ keV at 37 kpc, although the uncertainties are too large to make the drop significant. On the other hand, as shown in Section \ref{AGN}, the gas temperature within the central 20 kpc radius can be substantially lower.
The electron density (Fig. \ref{Tprofile} bottom left) was obtained directly from the deprojected normalization of 
the APEC component, assuming $n_{e} = 1.22n_{H}$, where $n_{e}$ and $n_{H}$ are the electron 
and proton densities. Based on the deprojected gas temperature and density, we calculated the 
pressure (Fig. \ref{Tprofile} top right) in each annulus as $P = nkT$ (where $n = 1.92n_{e}$) and 
the entropy (Fig. \ref{Tprofile} middle right) as $S = kT /n_{e}^{-2/3}$. The cooling time (shown 
in Fig. \ref{Tprofile} bottom right) is defined as $t_{cool} = 3P/[2n_{e}n_{H}\Lambda(T,Z)]$, 
where $\Lambda(T,Z)$ is the cooling function determined by the gas temperature and metallicity.
Assuming that the cluster was born at a redshift of 2.0, we estimated a cluster age of $\sim4.5$ Gyr. 
This gives a cooling radius of 71 kpc, if we define it as the radius where the cooling time equals 
the cluster age. We derive a bolometric luminosity of $1.5\times10^{44}$ erg s$^{-1}$ within the cooling radius.

Assuming hydrostatic equilibrium and spherical symmetry, we calculate the gravitational acceleration, $g$, as
\begin{equation}
	g=\frac{d\Phi}{dr} = -\frac{1}{\rho}\frac{dP}{dr},
\end{equation}
where $\Phi$ is the gravitational potential, $\rho=1.92n_{\rm e}\mu m_{\rm p}$ is the particle 
gas density, $\mu=0.60$ is the mean particle weight and	$m_{\rm p}$ is the proton mass. To obtain 
a smooth pressure profile, we fitted it with a generalized NFW model, where $P\propto r^{-3}$ at 
large radii and the inner slope is allowed to vary. We found the best-fitting inner slope of $0.58\pm0.11$. 
Once the gravitational acceleration is determined, we calculated the free fall time 
$t_{\rm ff}=\sqrt{\frac{2r}{g}}$ \citep[e.g.,][]{Gaspari12a}. As shown in Fig. \ref{tratio}, the 
minimum value of $t_{\rm cool}/t_{\rm ff}$ is $\sim30$, where the cooling time is $\sim1.7$ Gyr. 
Although this value is larger than the typical threshold ($\sim10-20$) for precipitation \citep{Sharma12,Gaspari12a,Voit15a},
it is possible that the threshold can be reached within the central $\sim20$ kpc.

We estimated the total gravitational mass assuming hydrostatic equilibrium for the hot gas in 
the gravitational potential. We already calculated the gravitational acceleration $g$, and the 
total gravitational mass is therefore, $M(r)= g(r)r^{2}/G$, where $G$ is the gravitational constant. 
The total gravitational mass profile is shown in Fig. \ref{mass}. The hydrostatic mass estimate is 
consistent with the previous mass estimate using strong lensing \citep[e.g.,][]{Comerford07}.

We measured X-ray temperature, $kT_{500}$, within $0.15-0.75 R_{500}$ ($\sim16\arcsec-80\arcsec$).
Using the surface brightness profile, after removing the PSF of the central AGN, we calculated the flux 
from the hot ICM in the central region (e.g., within $0.15R_{500}$) and outer region ($>0.75R_{500}$), 
and estimate a bolometric luminosity within $R_{500}$, $L_{X,500}$, of $2.5\pm0.4\times10^{44}$ erg s$^{-1}$. 
Based on the established $L-T$ relation, the estimated luminosity based on the fitted temperature, 
after correcting for the cosmological evolution term $E(z)$, is $\sim2.0\times10^{44}$ erg s$^{-1}$ 
from \citet{Giles16}, or $\sim4.2\times10^{44}$ erg s$^{-1}$ from \citet{Sun12}. 

\section{Discussion}
\subsection{The Central AGN}
\label{AGN}
\begin{table*}
\protect\caption{Spectral fits of the central AGN}
\begin{tabular}{|l|c|c|c|c|c|l|}
\hline
Model               & $\Gamma_{1}$$^a$            & $kT$$^b$        & $N_{\rm H, intr}$$^c$        &  $L_{0.5-2.0}$$^d$         & $L_{2-10}$$^e$  & [$C$,$C_{e}$,$C_{\sigma}$]$^f$   \\
&                         & (keV)                  & ($10^{22}$ cm$^{-2}$)  &  ($10^{43}$erg s$^{-1}$)      & ($10^{44}$erg s$^{-1}$)       &   \tabularnewline
\hline
tbabs*POW               & $1.62^{+0.02}_{-0.02}$  &                        &                        &                        & $5.17^{+0.07}_{-0.07}$ &  [166.9, 148.6, 17.3]\tabularnewline
tbabs*(ztbabs*POW+APEC) & $1.64^{+0.08}_{-0.08}$  & $1.05^{+0.20}_{-0.14}$ & $0.28^{+0.15}_{-0.08}$ & $3.77^{+0.82}_{-0.78}$ & $5.16^{+0.08}_{-0.08}$ & [148.0, 148.6, 17.3]  \tabularnewline
\hline
\end{tabular}
\begin{tablenotes}
 \item
	 $a$: the photon index for the power-law component;
	 $b$: the temperature for the thermal component (we assumed an abundance of 0.5 $Z_{\odot}$. 
	 The fits are not sensitive to the assumed abundance value.); 
	 $c$: the column density for the intrinsic absorption;
	 $d$: the rest-frame 0.5-2.0 keV luminosity for the thermal component;
	 $e$: the rest-frame 2-10 keV luminosity for the power-law component;
	 $f$: $C, C_{e}, C_{\sigma}$ are the fitted C-statistic, its expected value, and its standard 
	 deviation computed based on \citet{Kaastra17}. 68\% of the time the acceptable spectral models 
	 should have $-1< \frac{C-C_{e}}{C_{\sigma}} < 1$.
\end{tablenotes}
\label{tab_pointsource}
\end{table*}
To study the central AGN, we reprocessed the \chandra\ data by setting {\tt CHECK\_VF\_PHA} as ``no'', 
since it could remove good events in observations with bright point sources by setting to ``yes''.
We extracted spectra of the central X-ray source from each observation in a circular region with a 
radius of 2\arcsec, and the background spectra from an annulus from 2\arcsec\ to 4\arcsec.
The total net counts from four observations are $\sim8000$ over the 0.5-7.0 keV energy band.
The spectra from the four observations were fitted simultaneously between 0.5 and 7.0 keV.
The power-law photon index is $1.62\pm0.02$ from a single power-law model (Table \ref{tab_pointsource}), 
close to the canonical AGN value of 1.7. We attempted a model with an obscured AGN plus a soft 
component, which indeed improves the spectral fits (Table \ref{tab_pointsource}).
3C~220.1's AGN is obscured with a small intrinsic absorption column density of $\sim 0.3\times10^{22}$ 
cm$^{-2}$. The rest-frame 2-10 keV luminosity of the AGN is $\sim5.0\times10^{44}$ erg s$^{-1}$.
We found that the luminosity of the AGN is increased by $\sim5\%$ by setting {\tt CHECK\_VF\_PHA} 
to ``no''. Using a bolometric correction of 40 from \citet{Vasudevan07}, we estimate the bolometric 
luminosity of the central AGN to be $\sim2.0\times10^{46}$ erg s$^{-1}$. We considered the aperture 
correction factor for the central AGN luminosity, since the aperture (2\arcsec-4\arcsec) used as 
background still contains emission from the central AGN. However, based on the simulated PSF of the 
central point source, we found that the aperture correction factor is very small ($<1$\%). We 
checked the pileup fraction by using the model jdpileup \citep{Pileup} and found that 
the pileup fraction is $<2$\%.

We examined the 2-10 keV variability of the central point source. As shown in Table \ref{tab_obs}, 
the time span of the four observations is about half a year. Since AGN variabilities have different 
timescales, we also include the previous, short ACIS-S observation (Obs. ID 839) taken in 1999. 
We found the 2-10 keV flux changes by $\sim7$\% during the four recent observations, and the flux 
from new observations is $\sim30$\% higher than that from the old observation, based on our own AGN 
spectrum fitting from the old observation, and the previous study \citep{Belsole06}. In the new data, 
we also checked the possible X-ray emission associated with the radio features, e.g., radio jet and 
the hot spots. We chose a circular region for the hot spots and a box region for the radio jet to 
measure the observed X-ray emission as in \citet{Massaro15}. Background regions of the same shape and 
size, were chosen to avoid emission from other sources, with two boxes for the radio jet, and three 
circles for the hot spots. There is marginal X-ray emission detected from the eastern radio hotspots 
with a significance of $\sim1.6$ $\sigma$, and there is no X-ray emission detected from the western 
hotspot or from the radio jet. Assuming an absorbed power-law model with a photon index of 2.5, we estimated 
a 0.5-10 keV luminosity of $\sim3.7\times10^{41}$ and $\sim2.7\times10^{41}$ erg s$^{-1}$ 
for two eastern radio hot spots.

Based on the 5 GHz radio-core flux density \citep[25 mJy;][]{Giovannini88}, we calculated a radio 
luminosity at 5 GHz of $L_{\textrm{R}}\approx2\times10^{42}$ erg s$^{-1}$ assuming the spectra 
index of $-0.6$. Using the black hole fundamental plane \citep{Gultekin09}, i.e., the relation 
between the radio luminosity at 5 GHz, the X-ray 2-10 keV luminosity, and the black hole mass, 
we estimate a black hole mass of $\sim 1.3\times10^{9}$ M$_{\odot}$. Assuming a radiative 
efficiency $\epsilon=0.1$, we estimate an Eddington luminosity $L_{\rm Edd}$ of $1.6\times10^{47}$ erg s$^{-1}$, 
giving an Eddington ratio $L_{\rm bol}/L_{\rm Edd}$ of $\sim0.13$.

\subsection{Cavity Power}
\label{CavityPower}
\begin{figure}
\hbox{\hspace{5px}
\includegraphics[scale=0.5]{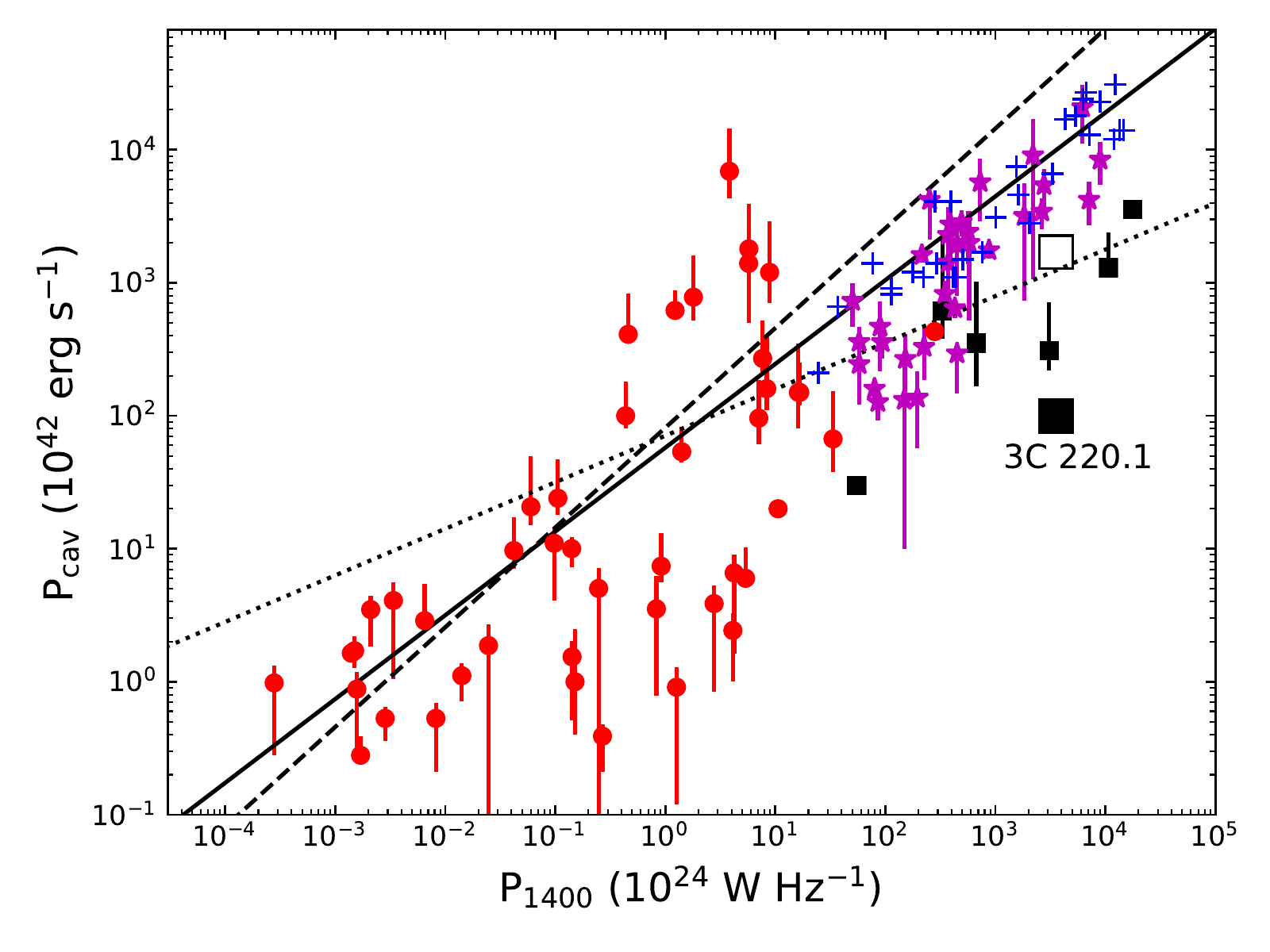}
}
\vspace{-0.3cm}
\caption{
	Cavity power versus radio power at 1.4 GHz for galaxy groups and clusters.
	Red circles denote FR-I sources from \citet{Birzan08}, \citet{Cavagnolo10}, and \citet{OSullivan11},
	with their best-fitting relations in dotted, dashed, and solid lines, respectively. For systems in common 
	between these papers we used the values from \citet{OSullivan11}.
	Black squares denote FR-II sources, i.e., Cygnus~A \citep{Rafferty06}, Hercules~A \citep{Rafferty06}, PKS~B2152-699 \citep{Worrall12}, 
	3C~295 \citep{Hlavacek-Larrondo12}, 3C~320 \citep{Vagshette19}, 3C~444 \citep{Vagshette17}, and 3C~220.1 (this work).
	Large solid square represents the cavity power of 3C 220.1 estimated from X-ray cavities, and 
	empty square represents the power estimated based on the radio lobes (both not corrected for projection).
	For comparison, we show FR-II sources with jet power estimated from other methods, i.e., based on the
	properties of hotspots \citep[blue plus;][]{Godfrey13}, and from FR-II lobes with X-ray IC emission
	detections \citep[magenta  star;][]{Ineson17}. See details in Section \ref{CavityPower}.
}
\vspace{-.5cm}
\label{P1400Pcav}
\end{figure}

In Fig. \ref{sb_in_out} (left panel), the potential cavities, defined by the surface brightness 
depressions, are close to the central nucleus and are smaller than the radio lobes. This is similar 
to some other high-redshift cool core systems, where the cavities are generally found in the central 
gas-rich regions \citep[e.g.,][]{Larrondo15}. This may suggest that cavities are created recently 
and have not buoyantly risen to the large radii. On the other hand, the actual cavities may be 
larger, since they could be obscured by the bright AGN at small radii and the enhanced X-ray emission 
from the lobes (discussed in Section \ref{Nonthermal}) and projection at large radii. 

We estimated a lower limit for the total jet power based on the enthalpy and buoyant rise time 
of the cavities \citep{Churazov01}, to allow a direct comparison with previous studies. 
We assumed the eastern and western cavities are ellipsoids as in Fig. \ref{sb_in_out}.
The projected distances of the eastern and western cavity centers to the nucleus are $\sim36.4$ 
kpc and $\sim35.0$ kpc. The major and minor axes are 14.6 kpc and 6.7 kpc for the eastern cavity, 
and 14.2 kpc and 7.2 kpc for the western cavity. The path length along the line of sight, 
calculated as the square root of the product of major and minor axes, are 9.9 kpc and 10.1 kpc, 
respectively. The estimated rise times of the bubbles to their present position are 
$8.8\times10^{7}$ yr and $8.3\times10^{7}$ yr. The total enthalpy ($H=4PV$, where $P$ is the 
azimuthally-averaged pressure at the radius of the center of the cavity) is calculated from 
the projected temperature and electron density profile. The total enthalpy and the cavity 
power for the two cavities are $2.7\times10^{59}$ erg and $\sim1.0\times10^{44}$ erg s$^{-1}$.
We note that the cavity power is slightly smaller than the cooling luminosity (section \ref{SpectralAnalysis}). 
Since the cavities here could be under-estimated, we also estimate the jet power assuming
the cavities are as large as the radio lobes. With a projected distance of $\sim65$ kpc to 
the nucleus, and major and minor axes of 40 and 20 kpc for both radio lobes, the total enthalpy 
and power for two cavities are $\sim3\times10^{60}$ erg and $\sim1.7\times10^{45}$ erg/s. We 
consider the actual cavity power in 3C~220.1 lies probably between these two estimates. 

If the cavity center does not lie on the plane of sky, we need to consider projection effect
on the estimated cavity power. Since the actual distance from the AGN to the center of cavity and the 
size of cavity would increase, the pressure at the center of cavity would be lower, while the
volume of cavity would increase, hence affecting the estimated total enthalpy. 
In addition, the estimated age of the cavity will increase. 
Based on the jet to counter-jet ratio, \citet{Worrall01} measured a maximum angle of
$\sim67$ degree of the jet to the line of sight for 3C~220.1, which can be assumed as the
upper limit for the angle of cavity to the line of sight.
We found that the estimated cavity power of 3C~220.1 increases with decreasing angle of the cavity
to the line of sight, e.g., the cavity power is $\sim20$\% larger for an angle of 60 degrees
and $\sim80$\% larger for 30 degrees.

For comparison, we estimated the jet power using other techniques. The jet power can be estimated from the observed parameters 
of the jet terminal hotspots as $Q_{hs}= Ac\frac{B_{eq}^{2}}{8\pi}\times g_{hs}$ \citep[e.g.,][]{Godfrey13}, 
where $A$ is the area of the jet hotspots, $c$ is the speed of light, $B_{eq}$ is the equipartition 
magnetic field strength, and $g_{hs}$ is the normalization factor, which can be empirically determined. 
Based on the 5 GHz radio flux at the eastern and western hotspots \citep{Hardcastle04}, we estimate 
an equipartition magnetic field of $\sim100$ $\mu$G for each hotspot \citep[e.g.,][]{Miley80}. 
From three hotspots, each with a radius of $\sim1.8$ kpc, and a $g_{hs}$ factor of 2, we 
estimate a total jet power of $\sim6.9\times10^{45}$ erg s$^{-1}$. The properties of hotspots 
are thought to be variable on timescales comparable to the light crossing time of a hotspot, 
and the jet power estimated from this technique could vary on a timescale much shorter than that 
used to estimate the cavity power. We note that jet power is about four times the estimated cavity 
power based on the radio lobes.

For a sample of powerful FR-II radio galaxies with X-ray inverse Compton (IC) emission detections, \citet{Ineson17} 
show that their radio lobes are typically dominated by relativistic electrons, and are in pressure balance
at lobe mid-point but over-pressured in their outer parts relative to the external medium \citep[e.g.,][]{Croston17}.
They presented another method to estimate the time-averaged jet power as $Q_{jet} \sim 2E_{i}/t \sim 2E_{i} \times v/l$, 
where $E_{i}$ is the internal energy of the lobe, $t$ is the age of the jet, $l$ is the lobe length, and $v$ 
is the advance speed of the lobe, estimated directly from lobe overpressuring.
With the enhanced X-ray emission possibly arising from IC emission (see Section \ref{Nonthermal}),
we estimated the internal pressure of 3C~220.1 and then its jet power.
We used the {\tt SYNCH}\footnote{https://github.com/mhardcastle/pysynch} code \citep{Hardcastle98}
and assumed a power law in electron energy with Lorentz factors between $\gamma_{min}=10$ and $\gamma_{max}=10^{5}$
with an energy index of 2.4. 
The eastern and western radio lobes are approximated as spheres with radii 
of $\sim34$ kpc\ and $\sim35$ kpc respectively, and their radio flux densities are from \citet{Croston05}.
We found that the observed magnetic field strengths of the eastern and western lobes are $\sim0.46B_{eq}$ and 
$\sim0.38B_{eq}$, respectively, close to the median value, $0.4B_{eq}$, found in \citet{Ineson17}.
We obtain internal pressures of $\sim5.0\times10^{-12}$ Pa and $\sim7.0\times10^{-12}$ Pa for
the eastern and western lobes, respectively, and the ratio to the external ICM pressures at the 
lobe mid-point are $\sim0.4$ and $\sim0.8$.
Therefore, based on the internal pressure, the pressure difference at the lobe-tip, and the gas temperature
(see Table \ref{tab_info}), we estimate a jet power of $\sim1.7\times10^{45}$ erg s$^{-1}$, which 
is consistent with the estimated cavity power based on the sizes of radio lobes.
We notice that the radio lobes appear slightly under-pressured with respect to the environment, and 3C~220.1
seems not follow the relation between the internal and external pressure presented in \citet{Croston17}, 
especially for the eastern lobe. However, the \citet{Croston17} relation is obtained from the studies of FR-II radio galaxies 
in galaxy groups. It has not been tested for the FR-IIs in massive galaxy clusters, where the pressure is much higher.
In addition, since 3C~220.1 resides in a denser environment, where the lobes are in direct contact with the
ICM, any entrained non-radiating particles will increase the internal pressure of the lobes. Alternatively,
the lobe pressure may be dominated by the kinetic and thermal energy of shocked gas, as for example in PKS B2152-699 \citep{Worrall12}.
Therefore, the jet power of 3C~220.1 could be under-estimated.

In Fig. \ref{P1400Pcav} we plot the radio power versus cavity power for FR-I sources and FR-II radio sources 
available from the literature \citep{Birzan08,Cavagnolo10,OSullivan11,Worrall12,Vagshette17,Liu19,Hlavacek-Larrondo12}. 
For comparison, we also show FR-II sources with their jet power estimated from two other techniques discussed above
\citep{Godfrey13,Ineson17}. We converted their radio luminosities at 151 MHz to the corresponding ones
at 1.4 GHz assuming a spectral index of 0.7. There are six common systems in two studies but the quoted radio luminosities and jet power are generally different, with the radio luminosity ratios from $0.2-2.8$ (a median value of 0.6) and the jet power ratios from
$0.8-9.6$ (a median value of 3.8).
As shown in Fig. \ref{P1400Pcav}, the jet power estimated from two different methods
are generally consistent for very luminous radio galaxies (e.g., P$_{1400}>\sim3\times10^{26}$ W Hz$^{-1}$). 
However, the power estimated from X-ray cavities are systematically smaller than the jet power estimated
from those two methods, especially
for the cavity power of 3C~220.1 determined from the inner pair of X-ray cavities.
We suggest the cavity power of 3C~220.1, or similar FR-II source, is under-estimated due to one or more of the following reasons:
1) A large amount of energy from the central SMBH is released as radiation power of the central AGN; 
2) Much of the energy of the central SMBH could be deposited in a shock, which either has dissipated, 
or is hard to detect in high redshift systems; 3) The sizes of cavities are under-estimated due to 
the complex structures, e.g., the bright central AGN and the enhancements in the lobes, projection and 
the limited data statistics.

We note that Fig. \ref{P1400Pcav} contains both FR-I and FR-II radio galaxies and the two populations
may not be fully comparable due to their different intrinsic properties and environment. To maintain pressure balance with the
external medium, their lobe internal composition is thought to be different, with FR-I lobes containing 
more protons \citep[e.g.,][]{Croston18}.
Therefore, the jet powers of the two populations are expected to be systematically different due to a high fraction
of non-radiating particles present in FR-I galaxies. In addition, the radio luminosity of an AGN depends on the
jet power and external pressure \citep[e.g.,][]{Hardcastle13} so a radio galaxy embedded in higher pressure environment would require a lower jet power to produce the same radio luminosity.

\subsection{Radiative and Kinetic Power}
\begin{figure}
\hbox{\hspace{5px}
\hspace{-0.5cm}
\includegraphics[height=17cm,width=8cm]{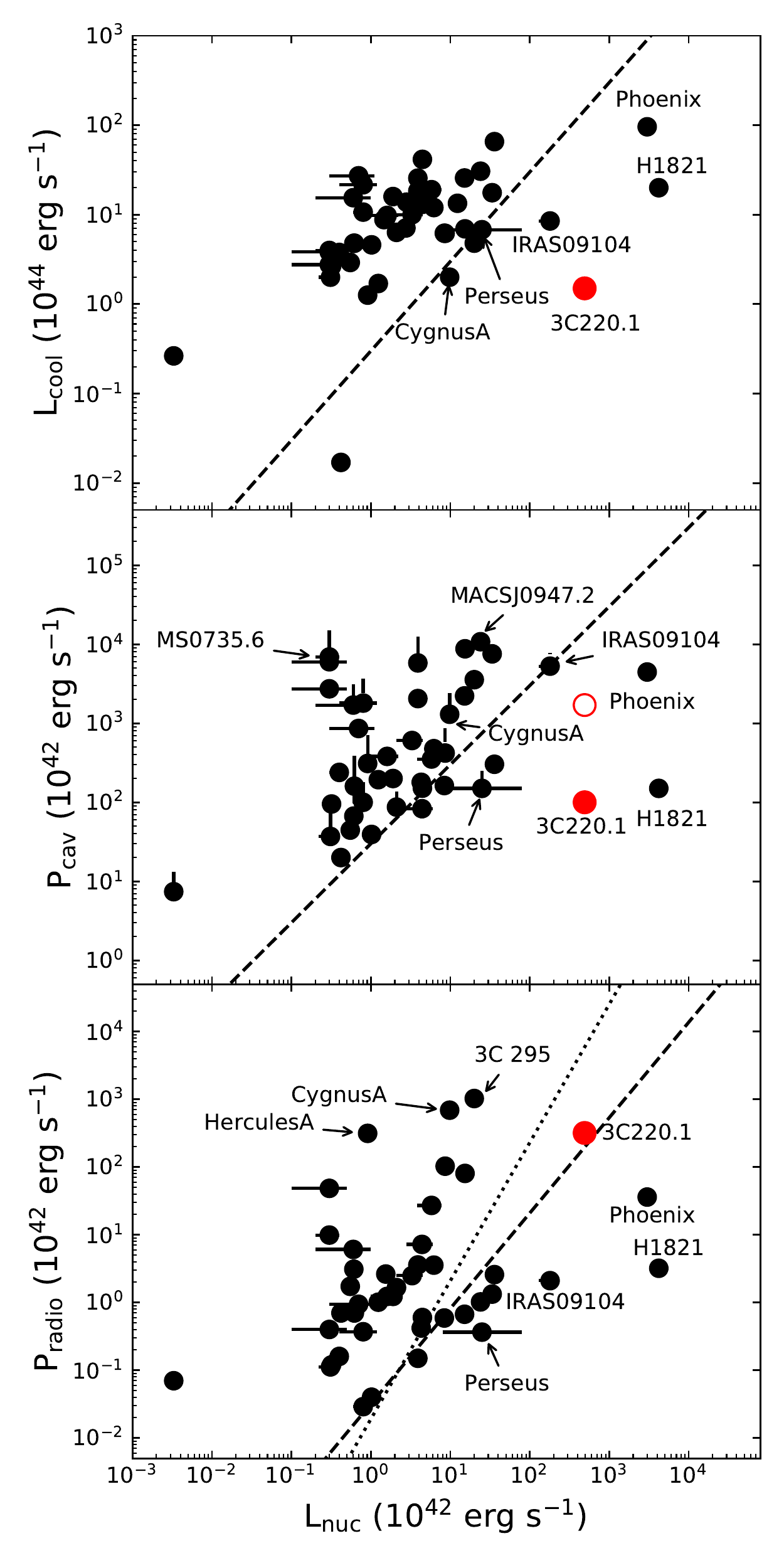}
}
\vspace{-0.5cm}
\caption{
	Top: The bolometric cooling luminosity versus nuclear 2-10 keV luminosity. The black dotted 
	line represents equal cooling and bolometric luminosity of central 
	AGN, assuming a bolometric correction factor of 30. 
	Middle: Cavity power versus nuclear 2-10 keV luminosity. The dashed line is the equal line 
	of cavity power and bolometric luminosity.
        Bottom: Total radio power (10 MHz to 10 GHz) versus nuclear 2-10 keV luminosity. The dashed 
	and dotted lines are the equal lines of the nuclear bolometric luminosity and the cavity power, 
	which are calculated based on the relation between the radio power and cavity power from 
	\citet{Birzan08} and \citet{OSullivan11}, respectively. The red circles are for 3C~220.1. The 
	empty red circle in the middle panel represents 3C~220.1, assuming that the cavities have the 
	same size as the radio lobes. H1821+643 data are from \citet{Russell10}. Phoenix Cluster data 
	are from \citet{McDonald12} and \citet{McDonald15}. IRAS 09104+4109 data are from \citet{Vignali11} 
	and \citet{OSullivan12}. Data of other black circles are from \citet{Larrondo11} and \citet{Larrondo13}. 
	Note that all the AGNs are located in the cluster BCGs.
}
\label{Lnuc}
\end{figure}

\begin{figure}
\hbox{\hspace{5px}
\includegraphics[scale=0.5]{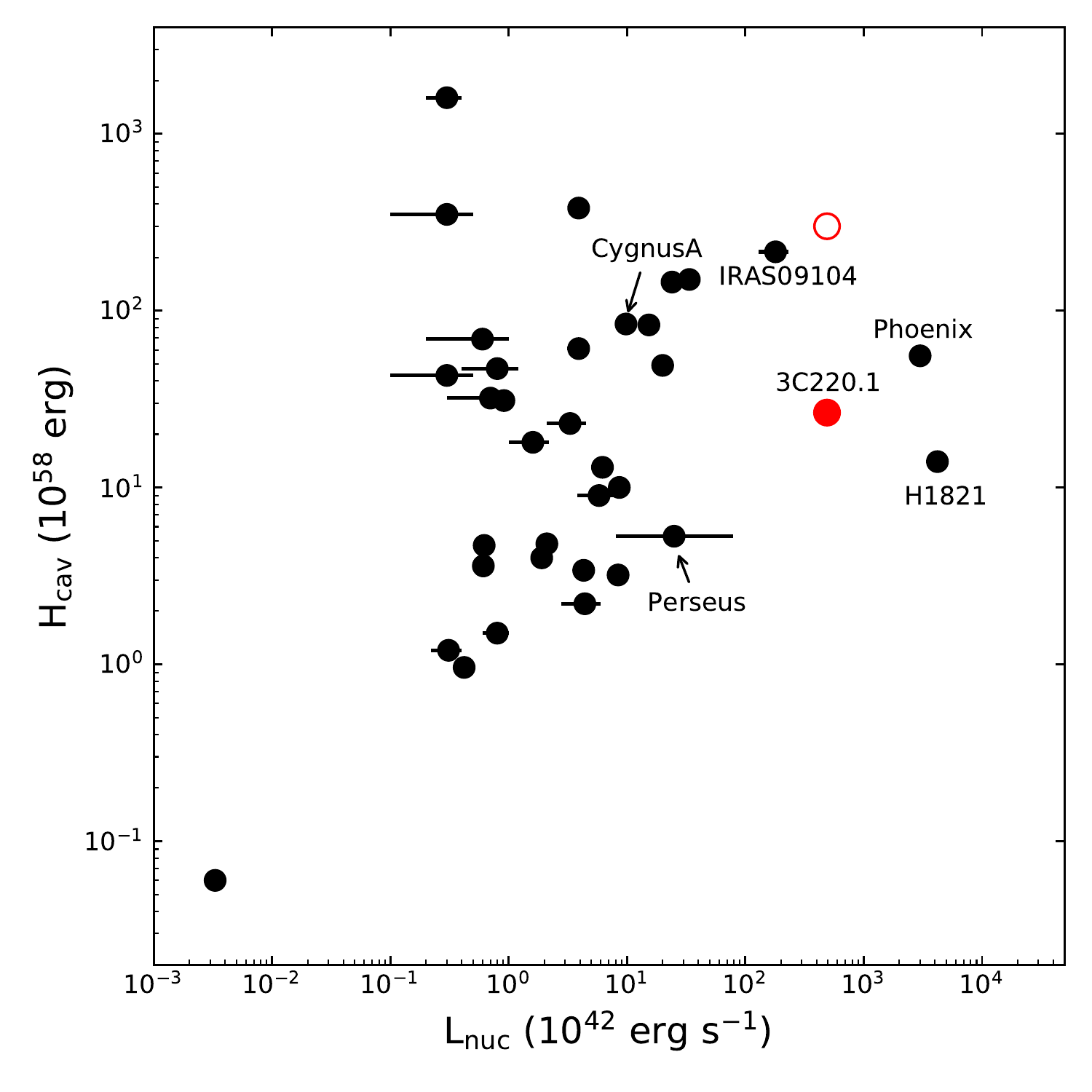}
}
\caption{
	Similar to Fig. \ref{Lnuc}, but for cavity enthalpy ($H_{\rm cav} = 4PV$) versus nuclear 2-10 keV 
	luminosity. The empty red circle represents 3C~220.1 assuming that the cavities have the same size 
	of the radio lobes. 
}
\label{LnucEcav}
\end{figure}

Fig. \ref{Lnuc} shows the comparison of the AGN nuclear 2-10 keV luminosity with the cooling luminosity, 
the cavity power, and the radio power for BCGs. Fig. \ref{LnucEcav} shows the cavity enthalpy versus 
the 2-10 keV nuclear luminosity. We included the samples from \citet{Larrondo11} and \citet{Larrondo13}, 
and three quasar systems, the Phoenix cluster \citep{McDonald12,McDonald15}, H1821+643 \citep{Russell10} 
and IRAS 09104+4109 \citep{Vignali11,OSullivan12}. The nuclear luminosities in the plot span more than 
five orders of magnitude. The properties of 3C~220.1 are similar to those of quasars. Both ``quasars'' 
and ``radio galaxies'' have luminous central AGNs, and energetic jets which extend outwards, inflate radio 
lobes, and create X-ray cavities. The nuclear luminosity of 3C~220.1 is the third largest in the plot, 
and is much greater than both the cooling luminosity and the cavity power. The ratio of nuclear 
luminosity (in the 2-10 keV band) to the cavity power for 3C~220.1 is $\sim5$, which is the second largest 
and only smaller than the ratio for H1821+643 ($\sim28$). Weak cavities could be missed and the cavity power 
in the middle panel can be taken as a lower limit. If we assume that the cavities in 3C~220.1 have the same 
size as the radio lobes, the cavity power would be almost 20 times larger, as shown by the open circle in 
the middle panel of Fig. \ref{Lnuc}. As an FR-II radio galaxy, 3C~220.1 produces a larger radio power than 
the three quasars shown.

A SMBH and its surrounding medium are expected to evolve through two stages \citep[e.g., see][and references therein]{Fabian12}. 
According to a standard paradigm, at early times the SMBH grows rapidly by accreting cooling gas at high rate,
and AGN feedback is radiation dominated in the so-called ``quasar-mode'' \citep[e.g.,][]{Springel05,Hopkins06,Sijacki06}. 
The SMBH produces enough power to suppress cooling, quenching star formation, and regulating the growth of the SMBH. As
the SMBH is starved of fuel and the accretion rate drops, the system evolves to the ``radio-mode'', where mechanical
dominated feedback regulates radiative cooling in the ICM in the form of bubbles and shocks 
\citep[e.g.,][]{McNamara07,Fabian12}. With a high accretion rate ($\sim$10\% of the Eddington rate), 3C~220.1 
is currently in quasar-mode. However, there is also evidence of mechanical heating, e.g., the radio jet 
and the X-ray cavities, and the mechanical power is able to compensate for the cooling losses of the ICM. 
Therefore, 3C~220.1 may be at the transition stage in which the SMBH has not evolved completely out of 
quasar-mode, but already produces sufficient power to suppress cooling.

Unlike most of the powerful FR-II galaxies at high redshift, 3C~220.1 lives in a dense environment. 
AGN activity with a high accretion rate can be triggered by a recent galaxy merger or interaction 
\citep[e.g.,][]{Ramos11,Pierce19}, as shown by the 
distorted morphology in \hst\ image (Fig. \ref{hst_image}) and the high star formation rate (section \ref{SFR}).
We can estimate the time elapsed since the last interaction and how long the AGN has been accreting in its current
mode through the spectral age. Assuming the magnetic field is constant and the particle injected have a
constant power-law energy spectrum, the spectral age is given by \citep[e.g.,][]{Jamrozy08}: 
\begin{equation}
	t=50.3[\nu(1+z)]^{-1/2}\frac{B^{1/2}}{B^{2}+B_{ic}^{2}} \textrm{Myr},
\end{equation}
where $B_{ic}=0.32(1+z)^{2}$, is the magnetic field strength in units of nT equivalent to the cosmic microwave background radiation,
$B$ is the magnetic field strength of the radio lobes in nT, $\nu$ is the spectral break frequency in GHz.
Although there is too little radio data to measure spectral curvature with sufficient angular resolution, the steep spectrum in lobe
suggests that the spectral break is below about 5 GHz. Using an equipartition magnetic field
of $\sim3$ nT, we estimate a lower limit to the age of $\sim3.2$ Myr. 

\subsection{BCG and Star Formation}
\label{SFR}
3C~220.1 has a 24 $\mu$m flux density of 2.2$\pm$0.2 mJy measured from {\em Spitzer} \citep{Cleary07}
and remains undetected at 70 $\mu$m, 100 $\mu$m and 160 $\mu$m from {\em Spitzer} and {\em Herschel}.
We assume an average spectral index of -0.3 between the rest-frame 15 $\mu$m
and 30 $\mu$m from the same work. The SFR of the BCG is then estimated to be
about 60 - 80 M$_{\odot}$/yr from the calibrations of \citet{Lee13} and \citet{Rieke09},
assuming a Kroupa IMF. This is consistent with the estimate from \cite{Westhues16} and comparable 
to the SFR of another high redshift cluster IRAS 09104+4109 ($z=0.44$) also with a luminous central 
AGN \citep{OSullivan12}. Substantial SF in the BCG of 3C~220.1 is also implied by its distorted 
morphology as seen with \hst\ (Fig. \ref{hst_image}). However, this estimate of SFR may be biased 
high if there is still a significant contribution from the AGN in the observed 24 $\mu$m band. 

\subsection{Deflected Radio Jet}
Fig. \ref{radio} (bottom right) shows a deflected radio jet (with an angle of $\sim14\arcdeg$) on 
the eastern side in the VLA 8.4 GHz map. It is almost straight to the bending point 
($\sim6\arcsec$ from the nucleus), and almost straight from that point to the second eastern 
hotspot \citep[e.g.,][]{Mullin06}, which may be where the jet partially de-collimates before being 
turned around the edge of the lobe into the first hotspot. This cannot be explained with the 
dentist's drill model \citep{Scheuer74}. In the dentist's drill model, when a hotspot is no longer 
being fed directly by the jet it will stop advancing and will expand to merge into the general 
radio lobe. This should happen rather fast, since the hotspot is a locus of high pressure excess, 
and the expansion time (greater than the radio fading time) is expected to be short compared with 
the jet propagation time from the core to the edge of the lobe. 

As described in \citet{Mullin06}, the flow continues from the jet to hotspots, and bends around to 
the south and back toward the core. 
The jet deflection point lies close to the edge (in projection) 
of the low-frequency radio lobe, so a natural explanation would be that the jet flow is encountering 
some angled interface between two different environments - passing into plasma dominated by the radio 
lobe. The structure can be interpreted in terms of low-density supersonic fluid flows with interactions 
causing the jets to change direction by a small angle until they slow down or encounter strong shocks. 

In \citet{Krause19}, the radio features of 3C~220.1 in the VLA 8.4 GHz image, such as jet curvature, 
jet at the edge of lobe, and the wide terminal hotspots on both sides, are considered as a result 
of jet precession. Jet precession usually produces curving jets, with S-symmetry if both jet and 
counter-jet are observed. Simple jet precession would have problems producing two such straight 
segments with a single bend. However, depending on the viewing angle, the jet can appear 
with mild curvature or as a straight jet.

\subsection{Enhanced X-ray Emission in the Radio Lobes}
\label{Nonthermal}
\begin{figure*}
\hbox{\hspace{10px}
\includegraphics[scale=0.42]{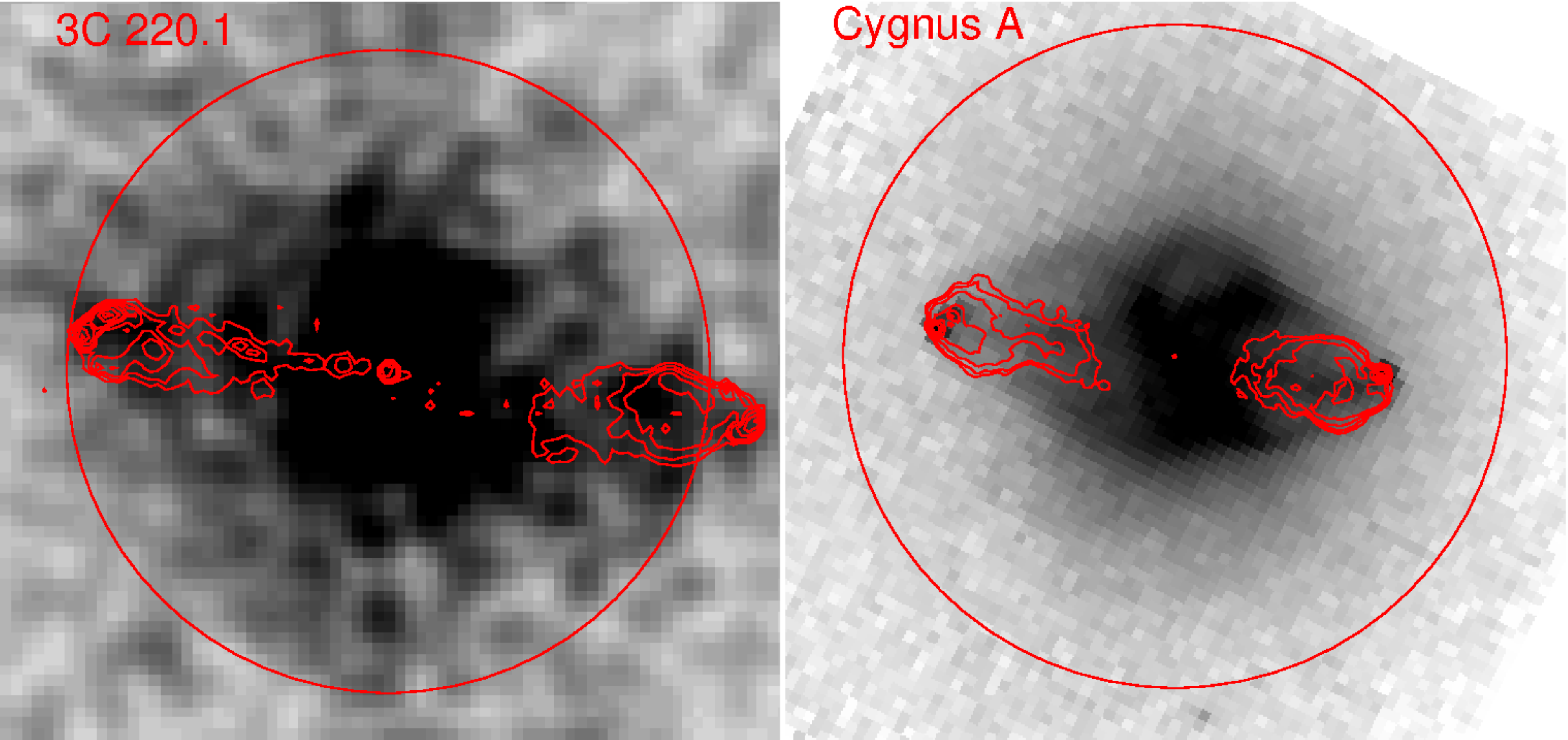}
}
\caption{
	Comparison of \chandra\ background-subtracted, exposure-corrected images of 3C~220.1 (left) 
	overlaid with VLA 8.4 GHz contours and Cygnus A (right) overlaid with VLA 1.5 GHz contours. 
	The Cygnus A image has been rotated to the same orientation as 3C~220.1, with the side with 
	a one-sided jet at small radii on the left as in 3C~220.1. The radii of the red circles are 
	100 kpc. To roughly simulate the X-ray image of Cygnus A at 3C~220.1's redshift, the X-ray
	image of Cygnus A has been re-binned by a factor of 6, and then re-scaled by the ratio of the squared 
	luminosity distances.
}
\label{Compare}
\end{figure*}

\begin{figure}
\hbox{\hspace{-10px}
\includegraphics[scale=0.32]{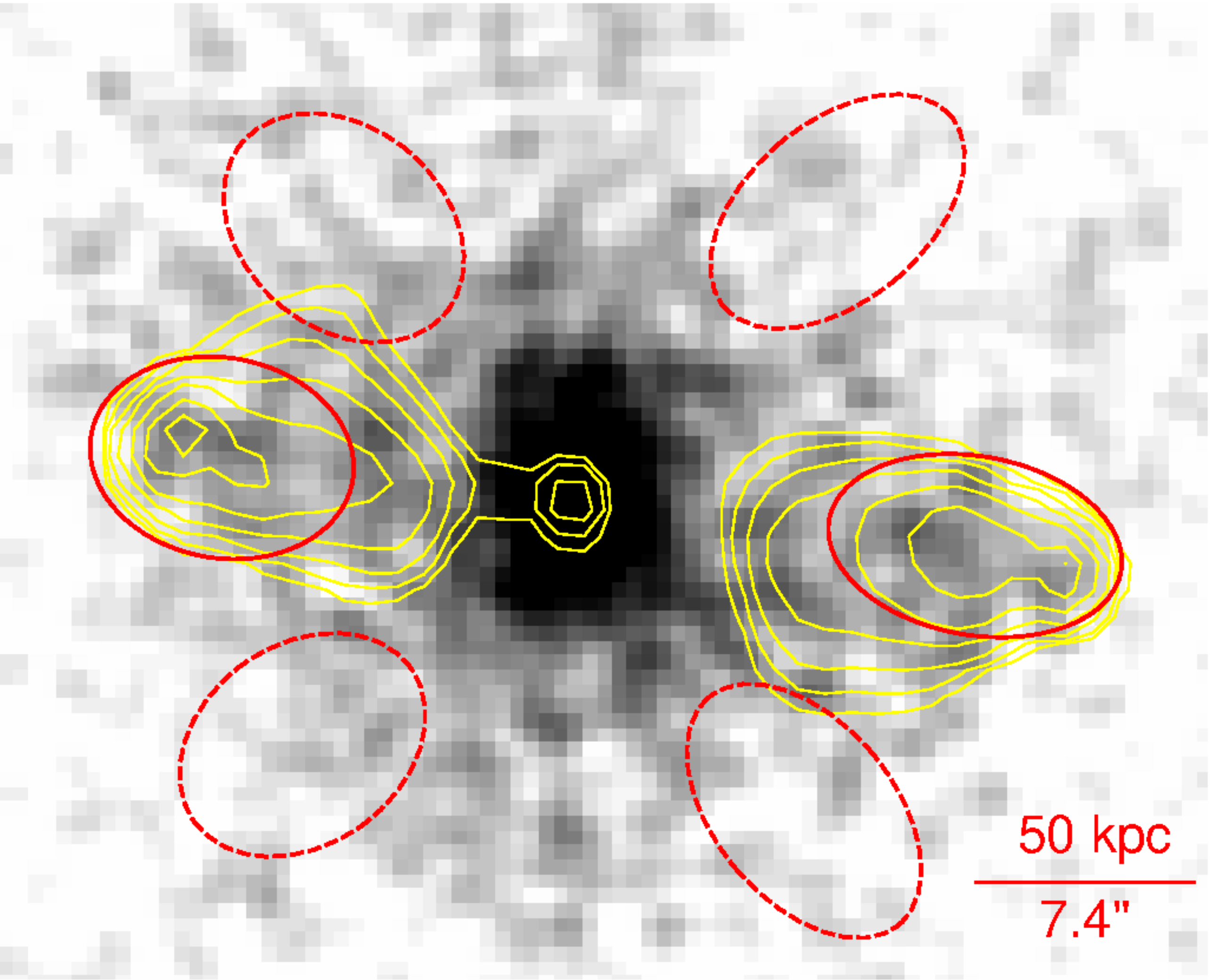}
}
\caption{
	Regions used to study the diffuse X-ray emission in the lobes of 3C~220.1. The radio contours at VLA
	1.5 GHz are shown in yellow. Solid ellipses are regions used for the eastern and western lobes; Dashed
	ellipses indicate the off-lobe regions.}
\label{IC_lobes}
\end{figure}

Powerful FR II radio sources show extended lobes of synchrotron-emitting plasma, as well as the jets 
which terminate in bright hotspots. The radio plasma displaces the hot ICM, forming X-ray cavities, 
and compresses it into a dense shell, which may be observed as a sharp bright edge in X-rays and is 
referred as the cocoon shock. One of the best-studied examples is the nearest powerful FR II source, 
Cygnus A \citep[e.g.,][]{Wilson06,Snios18}. As a powerful FR II source at a higher redshift, 3C~220.1 
shows structure similar to Cygnus A. In Fig. \ref{Compare} we show the \chandra\ images of 3C~220.1 
and Cygnus A side by side. Assuming that Cygnus A is observed at a similar redshift to 3C~220.1's, 
we re-binned the X-ray flux image of Cygnus A by a factor of 6 based on the pixel scale ratio at the two 
different redshifts (the actual pixel scale ratio is 6.19), then re-scaled it by the ratio of squared 
luminosity distances. The final image is then rotated to have similar tilted radio structures as 3C~220.1. 
Interesting features of the two systems are within similar physical scales ($\sim100$ kpc, the red circles 
in Fig. \ref{Compare}). Along the radio jet direction away from the central bright region, both systems 
show X-ray cavities in the radio lobes, and enhanced X-ray emission in the region toward the hotspots. 
Outside the radio lobes, there is a possible shock edge in 3C~220.1, which is difficult to confirm through 
the surface brightness profile due to the limited counts.

X-ray emission in the lobes can have contributions from inverse Compton (IC) emission, produced by photons 
being IC-scattered to X-ray energies by the radio-synchrotron emitting electrons. 
IC emission has been observed 
in many FR II radio lobes \citep[e.g.,][]{Isobe02,Bondi04,Hardcastle10,Croston05,Ineson17,DeVries18}. In order to check 
for IC X-ray emission in the lobes, we chose two elliptical regions for the lobes based on Fig. \ref{sb_in_out}, 
and two off-lobe regions for each lobe, as shown in Fig. \ref{IC_lobes}. We extracted the spectra 
from those regions from each observation. The spectra from the two off-lobe regions are combined with the 
ciao tool {\tt combine\_spectra}. The spectra from the lobe regions show clear excesses over those from 
off-lobe regions. Assuming the thermal emission from the hot ICM is the same in lobe- and off-lobe 
regions, we found that an extra power-law component with fixed index of 2.0 helps to fit the excess, 
suggesting the enhanced X-ray emission is of non-thermal origin. We obtained values for the power-law 
normalization at 1 keV of 0.5 nJy and 0.8 nJy for the eastern and western lobes, respectively, and 
luminosities at 2-10 keV of $\sim3\times10^{42}$ and $\sim5\times10^{42}$ erg s$^{-1}$. Considering 
the lobe regions in our study are smaller than those in \citet{Croston05}, our estimates are consistent 
with their results. However, we found a second thermal component with a high temperature 
(e.g., $\sim6$ keV) can also fit the excess. Due to the limited data statistics, we cannot make a 
definitive conclusion as to which model better describes the enhanced emission. 

\section{Summary}
3C~220.1 is an FR-II radio galaxy which is undergoing strong radiative feedback. We have analyzed
174 ks of \chandra\ observations centered on the BCG of this $kT \sim 4$ keV galaxy cluster and presented 
the properties of the hot gas related to AGN feedback. Our results are summarized as follows:
\begin{itemize}
	\item 
		The X-ray emission of the central AGN dominates the central $\sim2$\arcsec. The bolometric
		luminosity is estimated to be $2.0\times10^{46}$ erg s$^{-1}$, among the highest known for 
		a cluster BCG. Assuming a radiative efficiency of 0.1, we estimate an Eddington ratio of $\sim0.13$.
	\item 
		We found a pair of potential X-ray cavities to the eastern and western sides of the AGN 
		based on the surface brightness depressions. The total enthalpy and cavity power for the 
		cavities are estimated to be $2.7\times10^{59}$ erg and $\sim1.0\times10^{44}$ erg s$^{-1}$. 
		This cavity power is comparable to the cooling luminosity of the cluster. 
		If the X-ray cavities are assumed to be as large as the radio lobes, the total enthalpy and 
		cavity power become $\sim11$ and $\sim17$ times larger.
		Furthermore, if we consider the effect of projection, the cavity power could
		be double our estimate if the lobes lie close to the line of sight.
		Using two other techniques we estimated the jet power of $6.9\times10^{45}$ erg s$^{-1}$ and $1.7\times10^{45}$ erg s$^{-1}$
		for 3C~220.1 based on the hotspot properties and the properties
		of lobes with X-ray IC emission detections, respectively. The cavity power of 3C~220.1 and other
		FR-II sources, being systematically smaller than the jet power estimated from those two methods,
		could be well under-estimated.
	\item 
		A likely explanation of the X-ray enhancements in the radio lobes of 3C~220.1 is IC emission. 
		We estimate 2-10 keV luminosities of $\sim3\times10^{42}$ and $\sim5\times10^{42}$ erg s$^{-1}$ 
		from the eastern and western lobes, respectively. 
	\item 
		The properties of 3C~220.1 are similar to those of quasars. The ratio of X-ray nuclear luminosity 
		to cavity power is $\sim5$, which is the second largest known among BCGs after the source H1821+643. 
		We suggest that 3C~220.1 is at the transition stage from quasar-mode feedback to radio-mode feedback. 
\end{itemize}
\section{Acknowledgements}

We thank the referee for important comments and suggestions.
Support for this work was provided by the National Aeronautics and Space Administration through {\em Chandra} 
Award Number GO5-16115X, GO7-18118X and AR7-18016X issued by the {\em Chandra} X-ray Center, which is operated 
by the Smithsonian Astrophysical Observatory for and on behalf of the National Aeronautics Space Administration 
under contract NAS8-03060. We also acknowledge the support from NASA/EPSCoR grant NNX15AK29A and NSF grant 1714764.
This research has made use of data and/or software provided by the High Energy Astrophysics Science Archive Research
Center (HEASARC), which is a service of the Astrophysics Science Division at NASA/GSFC and the High Energy 
Astrophysics Division of the Smithsonian Astrophysical Observatory. The NRAO {\em VLA} Archive Survey image 
was produced as part of the NRAO {\em VLA} Archive Survey, (c) AUI/NRAO.

\bibliographystyle{mnras}
\bibliography{my}

\end{document}